\documentclass[a4paper,11pt]{article}
\usepackage{graphicx}
\usepackage{here}
\usepackage{amsmath,amssymb,amsthm}
\usepackage{mathrsfs}
\usepackage{ascmac}
\usepackage[sort, comma]{natbib}
\usepackage[subrefformat=parens]{subcaption}

\usepackage{url}
\usepackage{mathtools}
\usepackage{physics}
\mathtoolsset{showonlyrefs=true}

\newtheorem{theo}{Theorem}
\newtheorem{lem}{Lemma}
\newcommand{\ol}[1]{\overline{#1}}
\newcommand{\ul}[1]{\underline{#1}}
\newcommand{\cl}[1]{\mathcal{#1}}

\title{
{{\Large Continuous space core-periphery model\\with transport costs in differentiated agriculture}}
}

\author{Kensuke Ohtake\thanks{Center for General Education, Shinshu University, Matsumoto, Nagano 390-8621, Japan,
E-mail: k\_ohtake@shinshu-u.ac.jp}}
\date{March 10, 2025}

\begin{document}
\maketitle

\begin{abstract}
The core-periphery model with transport costs in differentiated agriculture is extended to continuous space. A homogeneous stationary solution is unstable but exhibits redispersion that it is stabilized by sufficiently low manufacturing transport costs or sufficiently strong preference for manufacturing variety. It is numerically observed that a solution starting from around the unstable homogeneous solution eventually forms a spike-like agglomeration. Furthermore, the redispersion also appears in the sense that the number of the spikes goes from decreasing to increasing as the manufacturing transport costs decrease. It is also observed that lower agricultural transport costs and stronger preference for agricultural variety promote agglomeration.
\end{abstract} 

\noindent
{\bf Keywords:\hspace{1mm}}
continuous racetrack economy;
economic agglomeration;
differentiated agricultural goods;
new economic geography;
self-organization;
transport costs

\vspace{1mm}

\section{Introduction}
The core-periphery model (CP model) proposed by \citet{Krug91} provides a sophisticated language based on general equilibrium theory for various fields involved in more or less market-economic and geographic phenomena, such as urban economics, regional economics, international economics, and regional science. The model has allowed these disciplines to be studied in a unified approach and has actually brought about the development of them in various directions, such as the formation of urban systems, population and industrial agglomeration, and international trade. 

In the most basic formulation of the CP model, agricultural goods are assumed to be undifferentiated, and, in addition, their transport is assumed to have no cost. Then, homogeneous stationary solution, where all regions are uniformly populated, is stable when manufacturing transport costs are sufficiently high, while it becomes unstable when manufacturing transport costs are sufficiently low. Meanwhile, \citet[Chapter 7]{FujiKrugVenab} introduce a model incorporating differentiated agriculture and its transport cost. This two-regional model has an interesting property that should be called {\it redispersion}. That is, the homogeneous stationary solution stabilizes not only when the manufacturing transport costs are sufficiently high but also when they are sufficiently low. Hence, the homogeneous stationary solution becomes unstable when the manufacturing transport costs are moderate.

In this paper, we consider the CP model with transport costs of differentiated agricultural goods in a continuous space, and investigate the behavior of the solution to the model in a one-dimensional periodic space. Firstly, the model of \citet[Chapter 7]{FujiKrugVenab} is extended to a continuous space. Secondly, the stability of the homogeneous stationary solution in which the population is uniformly distributed on the continuous space is investigated by using the Fourier analysis. Thirdly, time evolution of the solution is computed numerically.

To determine the stability of the homogeneous stationary solution, a small perturbation added to it is decomposed into the Fourier series of eigenfunctions each of which has an integer number of its own spatial frequency. For any eigenfunction having parameter area where its eigenvalue is positive, it is observed that the eigenvalue transitions from negative, positive, to negative as the manufacturing transport costs decrease. This means that the redispersion is also confirmed in the continuous space model. In addition, it is numerically suggested that a sufficient decrease in agricultural transport costs and a sufficient strengthening of  preference for agricultural variety promote agglomeration in the sense that the absolute values of frequencies of eigenfunctions having positive eigenvalues become smaller.

When the homogeneous stationary solution is unstable, the time-evolving solution eventually converges to an asymptotic distribution with some spikes. As manufacturing transport costs decrease, the average number of the spikes initially decreases and then increases. This implies that the redispersion is valid not only in the linear analysis around the homogeneous stationary solution, but also in terms of asymptotic distributions where nonlinearity is essentialy important. In addition, lower agricultural transport costs and stronger preference for agricultural variety are also observed to reduce the average number of the spikes. 

There are several related literature. \citet[Section 5]{OttThi2004} mentions transport costs of agricultural goods in a review of spatial economic models. \citet{PicaZeng05} consider a quasi-linear model with differentiated agriculture and its transport costs. In addition, \citet[Section 9.3]{ZenTaka} and \citet{Zeng2021} treat a footloose entrepreneur model with differentiated agriculture and its transport costs. In international trade theory, models including agriculture and its transport costs have also been devised. For example, \citet{TakaZeng2012mobilecapital} and \citet{TakaZeng2012} consider transport costs of homogeneous agriculture. \citet{ZeKiku09} consider those of differentiated agriculture. \citet{GaZo2018} study an impact of international trade on domestic economic geography based on a multi-regional CP model having transport costs of differentiated agriculture. While these studies consider discrete spatial models consisting of a finite number of regions (or countries), the present paper focuses on continuous spatial models.

The rest of the paper is organized as follows. Section 2 derives the model. Section 3 discusses the homogeneous stationary solution and the eigenvalue of the linearized problem around the homogeneous stationary solution. Section 4 examines how the eigenvalues depend on various parameters. Section 5 discusses the numerical results. Section 6 gives conclusions. Section 7 gives proofs and a numerical scheme omitted in the main part.

\section{The model}
In this section, we derive multi-regional and continuous space models with differentiated agriculture and its transport cost besed on the Dixit-Stiglitz framework.\footnote{\citet{DS77}}

\subsection{Modeling}\label{ss:modeling}

\subsubsection{Dixit-Stiglitz framework}

The economy is supposed to have a continuously infinite variety of manufactured goods and agricultural goods. Let $m(i)\geq 0$ and $a(j)\geq 0$ denote demands for the variety $i\in[0, V_M]$ of the manufactured goods and the variety $j\in[0, V_A]$ of the agricultural goods, respectively. The price of $i$-th variety of the manufactured goods is denoted by $p^{M}(i)$ and that of $j$-th variety of agricultural goods is denoted by $p^{A}(j)$. Given the prices of each variety of the goods and income $Y$, each consumer faces the following budget constraint.
\begin{equation}\label{bc}
\int_0^{V_M} p^{M}(i)m(i)di + \int_0^{V_A} {p^{A}}(i) a(i)di = Y.
\end{equation}
Each consumer is then supposed to maximize the following utility function
\begin{equation}\label{U}
M^\mu A^{1-\mu},~\mu\in(0,1),
\end{equation}
Here, $M$ and $A$ denote composite indices of the manufactured and the agricultural goods, respectively. The composite indices are given by
\begin{align}
&M = \left[\int_0^{V_M} m(i)^{\frac{\sigma-1}{\sigma}}di\right]^{\frac{\sigma}{\sigma-1}},\label{cmangood}\\
&A = \left[\int_0^{V_A} a(i)^{\frac{\eta-1}{\eta}}di\right]^{\frac{\eta}{\eta-1}},\label{cagrgood}
\end{align}
where $\sigma>1$ and $\eta>1$ and stand for the elasticity of substitution between any two varieties of the manufactured and that of the agricultural goods, respectively.  Following \citet[Chapter 4]{FujiKrugVenab}, we solve the maximizing problem \eqref{U} under \eqref{bc} in two steps: First, the minimum cost to achieve the partial utility $M$ and $A$ is calculated, and then the optimal expenditure on $M$ and $A$ under the budget constraint is determined.\footnote{For details of the two-stage budgeting, see  \citet[Chapter 5]{DeaMuell} or \citet[Chapters 2-4]{HAJGreen}}

As a result, we obtain the consumer demand functions for each variety of the manufactured and the agricultural goods as
\begin{align}
&m(i) = \mu Y {p^{M}(i)}^{-\sigma}{G^{M}}^{\sigma-1}, \label{mad} \\
&a(i) = (1-\mu)Y{p^{A}(i)}^{-\eta}{G^{A}}^{\eta-1}, \label{agd}
\end{align}
respectively. Here, $G^{M}$ and $G^{A}$ are the price index of the manufactured goods and the agricultural goods, respectively, defined by
\begin{align}
&G^{M}=\left[\int_0^{V_M} p^{M}(i)^{1-\sigma}di\right]^{\frac{1}{1-\sigma}}, \label{piM} \\
&G^{A}=\left[\int_0^{V_A} p^{A}(i)^{1-\eta}di\right]^{\frac{1}{1-\eta}}.\label{piA}
\end{align}

\subsubsection{Multi-regional extension}

Based on the framework described above, we model a two-sector spatial economy consisting of multiple regions. The modeling of the manufacturing sector has already been described in detail in \citet[Chapters 4, 5, and 6]{FujiKrugVenab}, but in order to make this paper self-contained, it is also given here, together with the modeling of the agricultural sector. 

For the moment, let us assume that the economy consists of $R\in\mathbb{N}$ discretely located regions indexed by $r$ or $s=1,2,\cdots,R$. There are two kinds of workers, manufacturing workers and agricultural workers, the former mobile in space, the latter immobile. Following \citet[Chapter 7]{FujiKrugVenab}, the total population of the manufacturing and agricultural workers are normalized to be $\mu$ and $1-\mu$, respectively. Let $\lambda_r$ and $\phi_r$ be the shares of region $r$ in the total population of the manufacturing and the agricultural workers, respectively, and let them satisfy that
\begin{align}
&\sum_{r=1}^{R} \lambda_r \equiv 1, \label{totm1}\\ 
&\sum_{r=1}^{R} \phi_r \equiv 1.\label{tota1}
\end{align}
Then, the manufacturing and the agricultural population in region $r$ is denoted by $\mu\lambda_r$ and $(1-\mu)\phi_r$, respectively. Let $w_r^{M}$ and $w_r^{A}$ denote the nominal wage of the manufacturing and agricultural workers, respectively. Then, the total income of region $r$ is given by
\begin{equation}\label{regincome}
Y_r = \mu w^{M}_r \lambda_r + (1-\mu) w^{A}_r\phi_r.
\end{equation}
One unit of any variety of the manufactured (resp. agricultural) goods produced in region $r$ have equal price $p^{M}_r$ (resp. $p^{A}_{r}$). Assume that the transportation of the goods incurs the iceberg transport costs.\footnote{The iceberg transport cost was formulated by \citet{Sa52}.} That is, to transport one unit of the manufactured (resp. agricultural) goods from $r$ to $s$, it is necessary to ship $T^{M}_{rs}\geq 1$ (resp. $T^{A}_{rs}\geq 1$) times as much manufactured (resp. agricultural) goods. It is  assumed that the cost is symmetric with respect to regions: $T_{rs}^M=T_{sr}^M$ and $T_{rs}^A=T_{sr}^A$. Then, the prices of each variety of the manufactured and the agricultural goods in region $s$ produced in region $r$ donoted by $p_{rs}^M$ and $p_{rs}^A$, respectively, are given by
\begin{align}
&p_{rs}^{M} = p^{M}_{r}T^{M}_{rs}, \label{prsmT} \\
&p_{rs}^{A} = p^{A}_{r}T^{A}_{rs}. \label{prsaT}
\end{align}
From \eqref{piM}, \eqref{piA}, \eqref{prsmT}, and \eqref{prsaT}, we see that the price index of each sector now takes different values in different regions as
\begin{align}
&G^{M}_r =
\left[\sum_{s=1}^R n^M_s {p^{M}_s}^{1-\sigma}{T^{M}_{sr}}^{1-\sigma}\right]^{\frac{1}{1-\sigma}},\label{piMr}\\
&G^{A}_r =
\left[\sum_{s=1}^R n^A_s {p^{A}_s}^{1-\eta}{T^{A}_{sr}}^{1-\eta}\right]^{\frac{1}{1-\eta}}.\label{piAr}
\end{align}

Let us consider producer behavior. In both sectors, a single firm is assumed to be engaged in the production of one variety of goods. The demand for each variety of the manufactured and the agricultural goods from a consumer having income $Y$ is given by \eqref{mad} and \eqref{agd}, respectively. Hence, the total demands from region $s$ for each variety of the manufactured and the agricultural goods produced in region $r$ are given as
\begin{align}
&q^M_{rs} = \mu Y_s {p^M_r}^{-\sigma}{T_{rs}^M}^{-\sigma}{G_s^M}^{\sigma-1},\label{DMsr} \\
&q^A_{rs} = (1-\mu)Y_s {p_r^{A}}^{-\eta}{T^{A}_{rs}}^{-\eta}{G^{A}_s}^{\eta-1},\label{DAsr} 
\end{align}
respectively. Multiplying \eqref{DMsr} and \eqref{DAsr} by $T^{M}_{rs}$ and $T^{A}_{rs}$, respectively, and adding up each over all regions yields the total sales of 
a manufacturing and an agricultural firm in region $r$ as
\begin{align}
&q^{M}_r = \sum_{s=1}^{R}q^M_{rs}T^{M}_{rs} =
\mu{p^{M}_r}^{-\sigma}\sum_{s=1}^{R}
 Y_s {T^{M}_{rs}}^{1-\sigma}{G^{M}_s}^{\sigma-1},\label{allDM}\\
&q^{A}_r = \sum_{s=1}^{R}q^A_{rs}T^{A}_{rs} =
(1-\mu){p^{A}_r}^{-\eta}\sum_{s=1}^{R}
 Y_s {T^{A}_{rs}}^{1-\eta}{G^{A}_s}^{\eta-1},\label{allDA}
\end{align}
respectively. It is assumed that the manufacturing sector is monopolistic competitive and that all firms have the same technology in which any firm requires the labor input $l^M$ given by
\begin{equation}\label{laborinput}
l^M = F + c^Mq^M
\end{equation}
to produce $q^M$. Here, $F$ and $c^M$ stand for a fixed input and a marginal input, respectively. Then, each firm in region $r$ maximizes the profit
\begin{equation}\label{profitr}
\pi^M_r = p^M_rq^M_r - w^M_r\left(F+c^Mq^M_r\right)
\end{equation}
under \eqref{allDM}.\footnote{As in \citet[p.51]{FujiKrugVenab}, the price indices $G_s^M$, $s=1,2,\cdots,R$ are considered given in the profit maximization.} Thus, we see that
\begin{equation}\label{optprice}
p_r^M = \frac{\sigma}{\sigma-1}c^Mw^M_r.
\end{equation}
Free entry results in zero equilibrium profit, therefore, we obtain the equilibrium output 
\begin{equation}\label{equiloutput}
{q_r^M}^* = \frac{F(\sigma-1)}{c^M}.
\end{equation}
from \eqref{profitr} and \eqref{optprice}. Then, from \eqref{laborinput} and \eqref{equiloutput}, the required labor input is
\begin{equation}\label{requiredl}
{l^M}^* = F\sigma.
\end{equation}
As a result, the number of varieties of manufactured goods in region $r$ is obtained by dividing the population of manufacturing workers there $\mu\lambda_r$ by the input per firm \eqref{requiredl} as
\begin{equation}\label{nrmulambdarFsigma}
n^M_r = \frac{\mu\lambda_r}{F\sigma}.
\end{equation}
The agricultural sector is perfect competitive. All firms have the same technology, however, there is no fixed input, and one unit of labor input is assumed to produce one unit of output. Therefore, the required labor input $l^A$ to produce $q^A$ of one variety of the agricultural goods is given by
\begin{equation}\label{laborinputA}
l^A = q^A
\end{equation}
All the agricultural firms are assumed to produce the same quantity
\begin{equation}\label{equiloutputA}
q^A \equiv {q^A}^*
\end{equation}
of agricultural goods, given exogenously. Then, the profit of an agricultural firm in region $r$ is
\begin{equation}\label{profitrA}
\pi^A_r = \left(p^A_r-w^A_r\right){q^A}^*.
\end{equation}
In this case, 
\begin{equation}\label{equilagrprice}
p^A_r = w^A_r
\end{equation}
holds in equilibrium.\footnote{It is a consequence of perfect competition under constant-return technology. See \citet[pp.192-193, pp.217-218]{HayashiMicro2021} for details.} The number of varieties of agricultural goods in region $r$ is given by
\begin{equation}\label{nAr1mmuphi}
n^A_r = \frac{(1-\mu)\phi_r}{{q^A}^*}.
\end{equation}

We now have equations that holds in market equilibrium. By applying \eqref{optprice} and \eqref{nrmulambdarFsigma} (resp. \eqref{equilagrprice} and \eqref{nAr1mmuphi}) to \eqref{piMr} (resp. \eqref{piAr}), we obtain price index equations as 
\begin{align}
&G^{M}_r = 
\left(\frac{\mu}{F\sigma}\right)^{\frac{1}{1-\sigma}}\frac{\sigma c^M}{\sigma-1}\left[\sum_{s=1}^R \lambda_s {w^{M}_s}^{1-\sigma}{T^{M}_{sr}}^{1-\sigma}\right]^{\frac{1}{1-\sigma}},\label{finalpiMr}\\
&G^{A}_r =
\left(\frac{1-\mu}{{q^A}^*}\right)^{\frac{1}{1-\eta}}\left[\sum_{s=1}^R \phi_s {w^{A}_s}^{1-\eta}{T^{A}_{sr}}^{1-\eta}\right]^{\frac{1}{1-\eta}}.\label{finalpiAr}
\end{align}
In market equilibrium, the demand \eqref{allDM} (resp. \eqref{allDA}) matches supply \eqref{equiloutput} (resp. \eqref{equiloutputA}) for each variety in each sector, thus by using \eqref{optprice} (resp. \eqref{equilagrprice}), we obtain nominal wage equations as
\begin{align}
&w^{M}_r = \frac{\sigma-1}{\sigma c^M}\left(\frac{F(\sigma-1)}{c^M\mu}\right)^{-\frac{1}{\sigma}}\left[\sum_{s=1}^{R}
 Y_s {G^{M}_s}^{\sigma-1} {T^{M}_{rs}}^{1-\sigma}\right]^{\frac{1}{\sigma}},\label{nomwageMr}\\
&w^{A}_r = \left(\frac{1-\mu}{{q^A}^*}\right)^{\frac{1}{\eta}}\left[\sum_{s=1}^{R}
 Y_s {G^{A}_s}^{\eta-1} {T^{A}_{rs}}^{1-\eta}\right]^{\frac{1}{\eta}}.\label{nomwageAr}
\end{align}

To simplify the equations to some extent, we normalize some units.\footnote{Note that in addition to the normalization of units below, we now choose the units of the number of workers so that the total number of the workers in each sector is $\mu$ and $1-\mu$, respectively.} Firstly, according to \citet[p.54]{FujiKrugVenab}, a unit of the manufactured goods output is chosen so that
\begin{equation}\label{normalized}
\frac{\sigma c^M}{\sigma-1}=1
\end{equation}
holds. Secondly, according to \citet[p.54]{FujiKrugVenab}, a unit of the number of the firms is chosen so that
\begin{equation}\label{normalizedF}
F = \frac{\mu}{\sigma}
\end{equation}
holds. Thirdly, we chose a unit of the agricultural goods output so that
\begin{equation}\label{normalizedqA}
{q^A}^* = 1-\mu
\end{equation}
holds.

The manufacturing workers' migration is driven by manufacturing real wages defined by discounting nominal wages by the price indices
\begin{equation}\label{realwageMr}
\omega^{M}_r =
w^{M}_r {G^{M}_r}^{-\mu} {G^{A}_r}^{\mu-1}.
\end{equation}
as in \citet[p.107]{FujiKrugVenab}. Then, we adopt the following ad-hoc dynamics as in \citet[p.62]{FujiKrugVenab} in which the manufacturing workers flow into regions with above-average real wage, while they flow out of regions with below-average real wage. The average real wage is defined by $\sum_{s=1}^{R}\omega_s\lambda_s$ and the dynamics is described by 
\begin{equation}\label{dyn}
\frac{d}{dt}\lambda_r = \gamma \left[\omega^M_r-\sum_{s=1}^{R}\omega^M_s\lambda_s\right]\lambda_r,
\end{equation}
where $t\geq 0$ is a time variable. Here, $\gamma>0$ represents the extent to which the manufacturing workers are sensitive to the gaps between the real wages in each region and the average real wage.

By specifying the time variable $t$ of the equations \eqref{regincome}, \eqref{finalpiMr}, \eqref{finalpiAr}, \eqref{nomwageMr}, \eqref{nomwageAr}, \eqref{realwageMr}, and \eqref{dyn} with the normalization \eqref{normalized}, \eqref{normalizedF}, and \eqref{normalizedqA}, we have the following system of nonlinear algebraic differential equations for $t\geq 0$ and $r=1,2,\cdots,R$.
\begin{equation}\label{dissys}
\left\{
\begin{aligned}
&Y_r(t) = \mu w^{M}_r(t) \lambda_r(t) + (1-\mu) w^{A}_r(t)\phi_r,\\
&G^{A}_r(t) =
\left[\sum_{s=1}^R \phi_s {w^{A}_s(t)}^{1-\eta}{T^{A}_{rs}}^{1-\eta}\right]^{\frac{1}{1-\eta}},\\
&w^{A}_r(t) = \left[\sum_{s=1}^{R}
 Y_s(t) {G^{A}_s(t)}^{\eta-1} {T^{A}_{rs}}^{1-\eta}\right]^{\frac{1}{\eta}},\\
&G^{M}_r(t) = 
\left[\sum_{s=1}^{R} \lambda_s(t) {w^{M}_s(t)}^{1-\sigma} {T^{M}_{rs}}^{1-\sigma} \right]^{\frac{1}{1-\sigma}},\\
&w^{M}_r(t) =
\left[\sum_{s=1}^{R} Y_s(t){G^{M}_s(t)}^{\sigma-1}{T^{M}_{rs}}^{1-\sigma}\right]^{\frac{1}{\sigma}},\\
&\omega^M_r(t) =
w^{M}_r(t) {G^{M}_r(t)}^{-\mu} {G^{A}_r(t)}^{\mu-1},\\
&\frac{d}{dt}\lambda_r(t) = \gamma \left[\omega^M_r(t)-\sum_{s=1}^{R}\omega^M_s(t)\lambda_s(t)\right]\lambda_r(t)
\end{aligned}
\right.
\end{equation}
with an initial condition $\lambda_r(0)\geq 0$ for $r=1,2,\cdots,R$ which satisfies $\sum_{r=1}^R \lambda_r(0)=1$.

\subsection{Continuous space model}
So far we have considered the model on the discrete regions $r=1,2,\cdots,R$, but now we consider the model on a continuous space. Let $K$ be a bounded set of a continuous space, typically a subset or a manifold in a multi-dimensional Euclidean space. We now consider the functions $\phi$, $Y$, $G^A$, $w^A$, $G^M$, $w^M$, $\omega$, and $\lambda$ as those on $[0,\infty)\times K$. Whereas in the discrete model, $(1-\mu)\phi_r$ and $\mu\lambda_r(t)$ represent the population of the workers in each sector in region $r$, in a continuous space model, $\mu\lambda(t, x)$ and $(1-\mu)\phi(x)$ represent the population {\it density} of the workers in each sector at region $x$, respectively. Then, the summation
\[
\sum_{s=1}^R f_{s}(t)
\]
is now replaced by the integration
\[
\int_K f(t, y) dy.
\]
For example, the second equation of the discrete space model \eqref{dissys} corresponds in a continuous space model to
\[
G^{A}(t,x) =
\left[\int_{K} \phi(y){w^{A}(t,y)}^{1-\eta} {T^{A}(x,y)}^{1-\eta}dy\right]^{\frac{1}{1-\eta}},
\]
where $T^A(x, y)$ stands for an iceberg-function corresponding to $T^A_{rs}$ in \eqref{dissys}. The iceberg-function satisfies that $T^A(x, y)\geq 1$, $T^A(x, y)=T^A(y, x)$, and $T^A(x, x)=1$ for any $x$ and $y$ in $K$. Another iceberg-function $T^M$ is also introduced in the same way.

Thus, we naturally obtain the following nonlinear system of integral-differential equations 
\begin{equation}\label{consys}
\left\{
\begin{aligned}
&Y(t,x)=\mu w^{M}(t, x)\lambda(t,x)+(1-\mu)w^{A}(t, x)\phi(x),\\
&G^{A}(t,x) =
\left[\int_{K} \phi(y){w^{A}(t,y)}^{1-\eta} {T^{A}(x,y)}^{1-\eta}dy\right]^{\frac{1}{1-\eta}},\\
&w^{A}(t,x) = \left[\int_{K}
 Y(t,y) {G^{A}(t,y)}^{\eta-1} {T^{A}(x,y)}^{1-\eta}dy\right]^{\frac{1}{\eta}},\\
&G^{M}(t,x) = 
\left[\int_{K} \lambda(t,y) {w^{M}(t,y)}^{1-\sigma} {T^{M}(x,y)}^{1-\sigma}dy\right]^{\frac{1}{1-\sigma}},\\
&w^{M}(t,x) =
\left[\int_{K} Y(t, y) {G^{M}(t,y)}^{\sigma-1}{T^{M}(x,y)}^{1-\sigma}dy\right]^{\frac{1}{\sigma}},\\
&\omega^M(t,x) =
w^{M}(t,x) {G^{M}(t,x)}^{-\mu} {G^{A}(t,x)}^{\mu-1},\\
&\frac{\partial}{\partial t}\lambda(t,x) = 
\gamma\left[\omega^M(t,x)-\int_{K} \omega^M(t,y)\lambda(t,y)dy\right]\lambda(t,x)
\end{aligned}
\right.
\end{equation}
with an initial condition $\lambda(0, x)\geq 0$ for all $x\in K$. Corresponding to \eqref{totm1} and \eqref{tota1}, the shares $\phi$ and $\lambda$ must satisfy that
\begin{equation}\label{totphi1}
\int_K\phi(x)dx\equiv 1,
\end{equation}
and
\begin{equation}\label{totlam1}
\int_K\lambda(t,x)dx\equiv 1
\end{equation}
for all $t\geq 0$.

\subsection{Racetrack setting}

In the following, we consider a so-called racetrack economy,\footnote{\citet[Chapter 6]{FujiKrugVenab}} that is, we take a circumference $S$ of a radius $\rho>0$ as a continuous space $K$. In addition, the iceberg-functions are also restricted to the forms 
\begin{equation}\label{TATM}
\begin{aligned}
&T^{A}(x,y)=e^{\tau^{A}d(x,y)},\\
&T^{M}(x,y)=e^{\tau^{M}d(x,y)},
\end{aligned}
\end{equation}
where $\tau^{A}\geq 0$ and $\tau^{M}\geq 0$ are cost parameters in each sector. Here, the distance function $d(x,y)$ is defined as the shorter distance between $x$ and $y$ along $S$. It is convenient to define 
\[
\begin{aligned}
&\alpha := \tau^{A}(\eta-1)\geq 0,\\
&\beta := \tau^{M}(\sigma-1)\geq 0,
\end{aligned}
\]
because the parameters $\tau_{A}$, $\tau_{M}$, $\eta$, and $\sigma$ often appear in this form.

We identify a function $f$ on $S$ with a corresponding periodic function on the interval $[-\pi,\pi]$ to perform specific calculations. In fact, this identification is possible in the following way. An element $x\in S$ corresponds one-to-one to an element $\theta\in[-\pi,\pi]$, i.e.,
\[
x = x(\theta).
\]
Therefore, for any function $f$ on $S$, there exists a corresponding function $\tilde{f}$ on $[-\pi, \pi]$ defined by
\begin{equation}
f(x)=f(x(\theta))=:\tilde{f}(\theta)
\end{equation}
which satisfies $\tilde{f}(-\pi)=\tilde{f}(\pi)$. In the following, we often write
\begin{equation}\label{abuse}
f(x) = f(\theta)
\end{equation}
by abuse of notation. Then, the integral of $f$ on $S$ is calculated by
\begin{equation}\label{racetrack integration}
\int_S f(x)dx = \int_{-\pi}^\pi f(\theta)\rho d\theta.
\end{equation}
The distance between $x=x(\theta)$ and $y=y(\theta^\prime)$ is calculated by
\[
d(x,y) = \rho \min\left\{|\theta-\theta^\prime|, 2\pi-|\theta-\theta^\prime|\right\}.
\]

\section{Stationary solution}
This section considers a homogeneous stationary solution and a linearized system of \eqref{consys} on $S$ with \eqref{TATM} around the stationary solution. In the following, we refer to ``\eqref{consys} on $S$ with \eqref{TATM}" simply as \eqref{consys} for brevity when no confusion can arise.

\subsection{Homogeneous stationary solution}
For the homogeneous agricultural population
\begin{equation}\label{homphi}
\phi(x)\equiv \ol{\phi} = \frac{1}{2\pi \rho},
\end{equation}
which satisfies \eqref{totphi1} due to \eqref{racetrack integration}, we find the homogeneous stationary solution such that all the functions $\lambda$, $Y$, $w^{A}$, $G^{A}$, $w^{M}$, $G^{M}$, and $\omega^M$ are spatially uniform. Let us denote homogeneous state of these functions by $\ol{\lambda}$, $\ol{Y}$, $\ol{w}^{A}$, $\ol{G}^{A}$, $\ol{w}^{M}$, $\ol{G}^{M}$, and $\ol{\omega}^M$, respectively. The value of this stationary solution can be obtained as the following theorem. See Subsection \ref{proof:thstsol} for the proof.
\begin{theo}\label{th: stationary solution}
Under \eqref{homphi}, the homogeneous states are given by
\begin{align}
&\ol{\lambda} = \frac{1}{2\pi \rho},\\
&\ol{Y} = \frac{\ol{w}}{2\pi\rho},\\
&\ol{w}^A = \ol{w}^M =: \ol{w}>0,\\
&\ol{G}^A = \left[\ol{\phi}{\ol{w}}^{1-\eta}E_\alpha\right]^{\frac{1}{1-\eta}},\\
&\ol{w}^M = \ol{w},\\
&\ol{G}^M = \left[\ol{\lambda}{\ol{w}}^{1-\sigma}E_\alpha\right]^{\frac{1}{1-\sigma}},\\
&\ol{\omega}^M = \ol{\lambda}^{\frac{\mu}{\sigma-1}}\ol{\phi}^{\frac{1-\mu}{\eta-1}}
E_\alpha^{\frac{1-\mu}{\eta-1}} E_\beta^{\frac{\mu}{\sigma-1}},
\end{align}
where $E_\alpha$ and $E_\beta$ are defined by
\[
\begin{aligned}
E_\alpha = \int_S e^{-\alpha|x-y|}dy = \frac{2(1-e^{-\alpha \rho\pi})}{\alpha},\\
E_\beta = \int_S e^{-\beta|x-y|}dy = \frac{2(1-e^{-\beta \rho\pi})}{\beta},
\end{aligned}
\]
respectively.
\end{theo}
\vspace{5mm}
\noindent
{\bf Remark}: 
It is interesting to note that the value of the homogeneous real wage $\ol{\omega}$ is uniquely determined even though the value of the homogeneous nominal wage $\ol{w}$ is not uniquely determined.

\subsection{Linearization}
Let $\varDelta Y(t,x)$, $\varDelta G^{A}(t,x)$, $\varDelta w^{A}(t,x)$, $\varDelta G^{M}(t,x)$, $\varDelta w^{M}(t,x)$, $\varDelta \omega^M(t,x)$, and $\varDelta \lambda(t,x)$ for $t\in[0,\infty)$ and $x\in S$ be small perturbations. Substituting $Y(t,x)=\ol{Y}+\varDelta Y(t,x)$, $G^{A}(t,x)=\ol{G}^{A}+\varDelta G^{A}(t,x)$, $w^{A}(t,x)=\ol{w}+\varDelta w^{A}(t,x)$, $G^{M}(t,x)=\ol{G}^{M}+\varDelta G^{M}(t,x)$, $w^{M}(t,x)=\ol{w}+\varDelta w^{M}(t,x)$, $\omega^M(t,x)=\ol{\omega}^M+\varDelta \omega(t,x)$, and $\lambda(t,x)=\ol{\lambda}+\varDelta \lambda(t,x)$ into the first six equations of \eqref{consys}, and ignoring higher order terms, we obtain the following linearized system
\begin{equation}\label{linsys}
\left\{
\begin{aligned}
&\varDelta Y(t,x) =
\mu\ol{w}\varDelta\lambda+\mu\ol{\lambda}\varDelta w^{M}(t,x)+(1-\mu)\ol{\phi}\varDelta w^{A}(t,x),\\
&\varDelta G^{A}(t,x) =
\frac{\ol{w}^{-\eta}{\ol{G}^{A}}^{\eta}}{2\pi \rho}
\int_S \varDelta w^{A}(t,y) e^{-\alpha|x-y|}dy,\\
&\varDelta w^{A}(t,x)
= \frac{\eta-1}{\eta}\ol{Y}\ol{w}^{1-\eta}{\ol{G}^{A}}^{\eta-2}
\int_S\varDelta G^{A}(t,y) e^{-\alpha|x-y|}dy\\
&\hspace{31mm} + \frac{{\ol{G}^{A}}^{\eta-1}\ol{w}^{1-\eta}}{\eta}
\int_S \varDelta Y(t,y) e^{-\alpha|x-y|}dy,\\
&\varDelta G^{M}(t,x)
=
{\ol{G}^{M}}^{\sigma}\ol{\lambda}\ol{w}^{-\sigma}
\int_S\varDelta w^{M}(t,y) e^{-\beta|x-y|}dy\\
&\hspace{34mm} + \frac{{\ol{G}^{M}}^{\sigma}}{1-\sigma}\ol{w}^{1-\sigma}
\int_S \varDelta\lambda(t,y) e^{-\beta|x-y|}dy,\\
&\varDelta w^{M}(t,x) =
\frac{\sigma-1}{\sigma}\ol{w}^{1-\sigma}\ol{Y}{\ol{G}^{M}}^{\sigma-2}
\int_S \varDelta G^{M}(t,y) e^{-\beta|x-y|}dy\\
&\hspace{33mm} + \frac{\ol{w}^{1-\sigma}}{\sigma}{\ol{G}^{M}}^{\sigma-1}
\int_S\varDelta Y(t,y) e^{-\beta|x-y|}dy,\\
&\varDelta\omega^M(t,x) = 
-\mu\ol{\omega}^M{\ol{G}^{M}}^{-1} \varDelta G^{M}(t,x)
-(1-\mu)\ol{\omega}^{M}{\ol{G}^{A}}^{-1} \varDelta G^{A}(t,x)\\
&\hspace{69mm} +\ol{\omega}^{M}\ol{w}^{-1} \varDelta w^{M}(t,x).
\end{aligned}
\right.
\end{equation}

We identify a perturbation function $\varDelta f(t, x)$ with $\varDelta f(t, \theta)$ as in \eqref{abuse} and define the Fourier expansion of  $\varDelta f$ with respect to $\theta \in [-\pi, \pi]$ as 
\[
\varDelta f(t, \theta) = \frac{1}{2\pi}\sum_{n=0,\pm1,\pm2,\cdots} \hat{f}_n(t) e^{in\theta}
\]
where $\hat{f}_n,~n=0,\pm1,\pm2,\cdots$ are the Fourier coefficients defined by
\[
\hat{f}_n(t) = \int_{-\pi}^\pi \varDelta f(t, \theta^\prime)e^{-in\theta^\prime}d\theta^\prime.
\]
By expanding the perturbatins in \eqref{linsys} into the Fourier serieses, we obtain the following equations of the Fourier coefficients
\begin{equation}\label{Fsys}
\left\{
\begin{aligned}
&\hat{Y}_n = 
\mu\ol{w}\hat{\lambda}_n + \mu\ol{\lambda}\hat{w}^{M}_n
+ (1-\mu)\ol{\phi}\hat{w}^{A}_n,\\
&\hat{G}^{A}_n = 
\ol{w}^{-1}\ol{G}^{A} H^\alpha_n \hat{w}^{A}_n,\\
&\hat{w}^{A}_n =
\frac{\eta-1}{\eta}\ol{w}{\ol{G}^{A}}^{-1}H^\alpha_n \hat{G}^{A}_n
+ \frac{2\pi \rho}{\eta}H^\alpha_n\hat{Y}_n,\\
&\hat{G}^{M}_n
= \frac{\ol{G}^{M}}{\ol{w}}H^\beta_n \hat{w}^{M}_n
+ \frac{2\pi \rho}{1-\sigma}\ol{G}^{M}H^\beta_n\hat{\lambda}_n,\\
&\hat{w}_n^{M}=\frac{\sigma-1}{\sigma}\frac{\ol{w}}{\ol{G}^{M}}H^\beta_n \hat{G}^{M}_n
+ \frac{2\pi \rho}{\sigma}H^\beta_n \hat{Y}_n,\\
&\hat{\omega}_n^{M} =
-\mu\ol{\omega}^{M}{\ol{G}^{M}}^{-1} \hat{G}^{M}_n
-(1-\mu)\ol{\omega}^{M}{\ol{G}^{A}}^{-1} \hat{G}^{A}_n
+\ol{\omega}^{M}\ol{w}^{-1} \hat{w}^{M}_n
\end{aligned}
\right.
\end{equation}
for $n=\pm 1,\pm 2,\cdots$. Here, $H^\alpha_n$ and $H^\beta_n$ are defined as
\begin{align}
&H^\alpha_n :=
\frac{\alpha^2\rho^2\left(1-(-1)^{|n|}e^{-\alpha \rho\pi}\right)}{(n^2+\alpha^2\rho^2)(1-e^{-\alpha \rho\pi})},\label{def:Halphan}\\
&H^\beta_n :=
\frac{\beta^2\rho^2\left(1-(-1)^{|n|}e^{-\beta \rho\pi}\right)}{(n^2+\beta^2\rho^2)(1-e^{-\beta \rho\pi})},\label{def:Hbetan}
\end{align}
respectively. These variables have the following properties. See Subsection \ref{proof:Hs} for the proof.
\begin{lem}\label{Hs}
$H^{\alpha}_n$ and $H^\beta_n$ are monotonically increasing for $\alpha \rho\geq 0$ and $\beta \rho\geq 0$, respectively. Furthermore, 
\begin{align}
&\lim_{\alpha \rho \to 0} H^\alpha_n = 0,\label{Halp0to0}\\
&\lim_{\alpha \rho \to \infty} H^\alpha_n = 1,\\
&\lim_{\beta \rho \to 0} H^\beta_n = 0,\label{Hbeta0to0}\\
&\lim_{\beta \rho \to \infty} H^\beta_n = 1.
\end{align}
\end{lem}

Additionally, let us define the following variables \footnote{Here, \eqref{b}, \eqref{D}, and \eqref{B} correspond to (7 A.8) in \citet[p.112]{FujiKrugVenab}, (7 A.14) in \citet[p.113]{FujiKrugVenab} and (7 A.23) in \citet[p.114]{FujiKrugVenab}, respectively. To be precise, \eqref{D} corresponds to (7 A.14) multiplied by $1-\sigma$.}
\begin{align}
&b:=
1-\frac{(1-\mu)H^\alpha_n}{\eta-(\eta-1){H^\alpha_n}^2},\label{b}\\
&D:= \sigma-\frac{\mu}{b}H^\beta_n-(\sigma-1){H^\beta_n}^2,\label{D}\\
&B:=\frac{\mu\sigma(\sigma-1)(1-b)H^\alpha_n}{b}.\label{B}
\end{align}
and the quadratic function
\begin{equation}\label{QH}
Q(H) := 
\left\{-\frac{\sigma(\mu^2+b)}{b}+1+B\right\}{H}^2
+\frac{\mu(\sigma(b+1)-1)}{b}H - B.
\end{equation}
The following lemmas hold as for $b$, $D$, and $B$. See Subsections to \ref{proof:delomega} for the proofs.
\begin{lem}\label{mub1}
$b$ defined by \eqref{b} is monotonically decreasing as a function of $H^\alpha_n$ and satisfies that $\mu<b<1$ for $\alpha>0$, and that $b=1$ for $\alpha=0$.
\end{lem}

\begin{lem}\label{Dgt0}
$D$ defined by \eqref{D} satisfies that $D>0$.
\end{lem}

\begin{lem}\label{Bgt0}
$B$ defined by \eqref{B} satisfies that $B>0$ for $\alpha>0$, and that $B=0$ for $\alpha=0$.
\end{lem}
By solving \eqref{Fsys}, we obtain
\begin{equation}\label{hatomegaOmeganhatlambda}
\hat{\omega}^{M}_n = \Omega_n \hat{\lambda}_n.
\end{equation}
Here,
\begin{equation}\label{omhat}
\begin{aligned}
\Omega_n &=
\frac{2\pi \rho\ol{\omega}^{M}}{(\sigma-1) D}  Q(H^\beta_n).
\end{aligned}
\end{equation}
The next lemma is useful. See Subsection \ref{proof:delomega} for the proof.
\begin{lem}\label{lem: delomega}
For the perturbation $\varDelta\omega^{M}$ in \eqref{linsys},
\begin{equation}\label{intdeltaomega0}
\int_S\varDelta\omega^{M}(t, x)dx = 0.
\end{equation}
\end{lem}
Thanks to \eqref{intdeltaomega0}, linearizing the last equation of \eqref{consys} yields
\begin{equation}\label{lindiff}
\frac{\partial}{\partial t}\varDelta\lambda(t,x) = \gamma\ol{\lambda}\varDelta\omega^{M}(t,x).
\end{equation}
By expanding $\varDelta \lambda$ and $\varDelta\omega$ of \eqref{lindiff} into the Fourier serieses and applying \eqref{hatomegaOmeganhatlambda}, we obtain the differential equations 
\begin{equation}\label{Fourierdiff}
\frac{d}{dt}\hat{\lambda}_n(t) = 
\gamma\ol{\lambda}\Omega_n \hat{\lambda}_n(t)
\end{equation}
for $n=\pm1, \pm2, \cdots$. It is easy to solve \eqref{Fourierdiff} as
\begin{equation}\label{expsolution}
\hat{\lambda}_n(t) = \hat{\lambda}_n(0)e^{\gamma\ol{\lambda}\Omega_n}
\end{equation}
for $n=\pm1, \pm2, \cdots$. The solution \eqref{expsolution} means that 1) if $\gamma\ol{\lambda}\Omega_n>0$, the absolute value of the amplitude of the $n$-th mode $\hat{\lambda}_ne^{in\theta}$ amplify to infinity, i.e., $n$ population agglomerations are formed at least initially,\footnote{Of course, the linear approximation is valid only when the amplitude of the perturbation is small, so the amplitude does not actually grow infinitely.} and 2) if $\gamma\ol{\lambda}\Omega_n<0$, the absolute amplitude of the $n$-th mode converges to $0$ and the spatial heterogeneity is dispersed and returns to the homogeneous state. The equation \eqref{Fourierdiff} states that 
\begin{equation}\label{ev}
\gamma\ol{\lambda}\Omega_n
\end{equation}
is the eigenvalue of the system \eqref{linsys}, and the $n$-th mode $\hat{\lambda}_ne^{in\theta}$ is the corresponding eigen function. Thus, the homogeneous stationary solution given in Theorem \ref{th: stationary solution} is unstable when the eigenvalue \eqref{ev} is positive for at least one frequency $n$, and it is stable when the eigenvalue \eqref{ev} is negative for all frequencies. Let us call a mode with positive (resp. negative) eigenvalue an {\it unstable} (resp. {\it stable}) {\it eigenfunction}. It is obvious from \eqref{ev} that the sign of the eigenvalue is that of $\Omega_n$.

\vspace{5mm}
\noindent
{\bf Remark}: 
By replacing $\zeta$ and $Z$ in \citet[Chapter 7]{FujiKrugVenab} with $H^\alpha_n$ and $H^\beta_n$, respectively, we see that Eqs. (7 A.1)-(7 A.3), (7 A.5), (7 A.6), and (7 A.21) in \citet[pp.111-112, p.114]{FujiKrugVenab} are practically equivalent to the quations of \eqref{Fsys}. Then, \eqref{omhat} corresponds to (7 A.22) in \citet[p.114]{FujiKrugVenab}. As can been seen, the mathematical results in this paper are a natural extension of the results obtained by \citet[Chapter 7]{FujiKrugVenab} to infinite-dimensional spaces.

\subsection{Eigenvalues in some extreme situations}\label{subsec:evext}
When a so-called ``no black hole condition" holds, the eigenfunction can be stabilized for sufficiently high manufacturing transport costs (or sufficiently weak preference for manufacturing variety). The no black hole condition same as \citet[p.59]{FujiKrugVenab} is given by the following theorem. See Subsection \ref{proofNBH} for the proof.
\begin{theo}\label{Th:NBH}
If 
\begin{equation}\label{nbh}
\frac{\sigma-1}{\sigma} > \mu
\end{equation}
holds, then $\lim_{\beta\to\infty}\Omega_n < 0$ for $n=\pm 1,\pm 2,\cdots$.
\end{theo}

Meanwhile, for sufficiently low manufacturing transport costs or sufficiently strong preference for manufacturing variety, the eigenvalue \eqref{ev} also takes negative value as the following theorem states.\footnote{In Theorem \ref{th:betatozero}, if $\alpha$ converges to zero, the limit value  \eqref{tm0ev} also converges to zero because $\lim_{\alpha\to 0} B=0$ from Lemma \ref{Bgt0}.} This means so-called redispersion in which the homogeneous stationary solution is stabilized by sufficiently low manufacturing transport costs or sufficiently strong preference for manufacturing variety. See Subsection \ref{betatozero} for the proof.

\begin{theo}\label{th:betatozero}
For any value of $\alpha>0$,  
\begin{equation}\label{tm0ev}
\lim_{\beta\to 0}\Omega_n = -\frac{2\pi\rho\ol{\omega}}{(\sigma-1)\sigma}B < 0.
\end{equation}
\end{theo}

The next theorem claims that there is always a positive eigenvalue for a sufficiently large $|n|$. See Subsection \ref{proofpev} for the proof.
\begin{theo}\label{th:pev}
For any $\beta>0$, there exists $N>0$ such that for $|n|>N$,
\[
\Omega_n > 0.
\]
\end{theo}

Meanwhile, when the absolute value of frequencies is sufficiently large, the eigenvalue converges to zero. See Subsection \ref{Omegatozero} for the proof.
\begin{theo}\label{th:Omegatozero}
For any $\alpha>0$ and $\beta>0$, 
\[
\lim_{|n|\to\infty}\Omega_n =0.
\]
\end{theo}

\section{Eigenvalue in detail}
This section discusses how the eigenvalue \eqref{ev} depends on the frequency of the eigenfunction and the parameters of the model. 

\subsection{Critical points}\label{subsec:evasfunc}
\begin{figure}[H]
\begin{subfigure}[H]{0.5\columnwidth}
\centering
\includegraphics[width=\columnwidth]{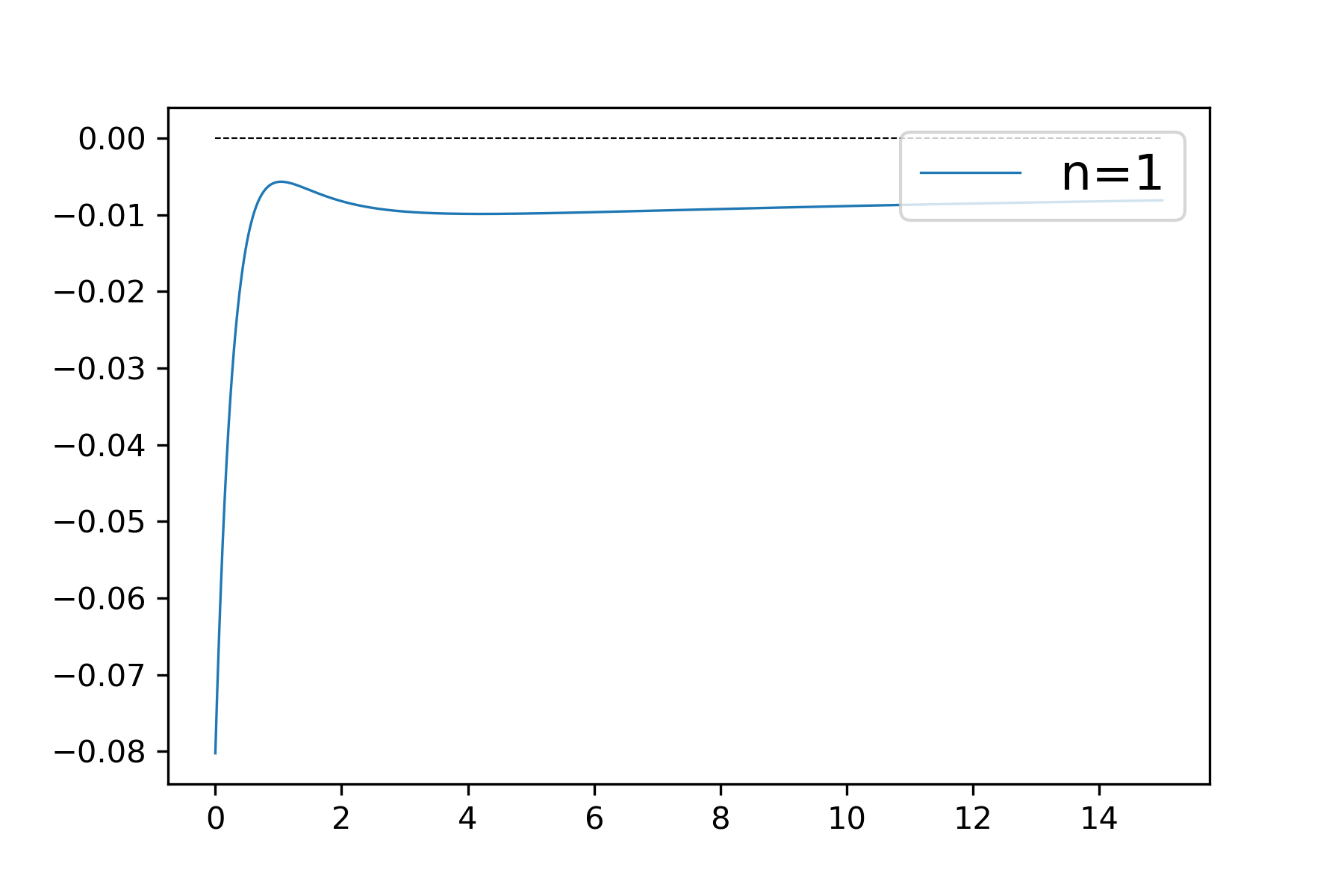}
\caption{Having no unstable area}\label{eigs0}
\end{subfigure}
\begin{subfigure}[H]{0.5\columnwidth}
\centering
\includegraphics[width=\columnwidth]{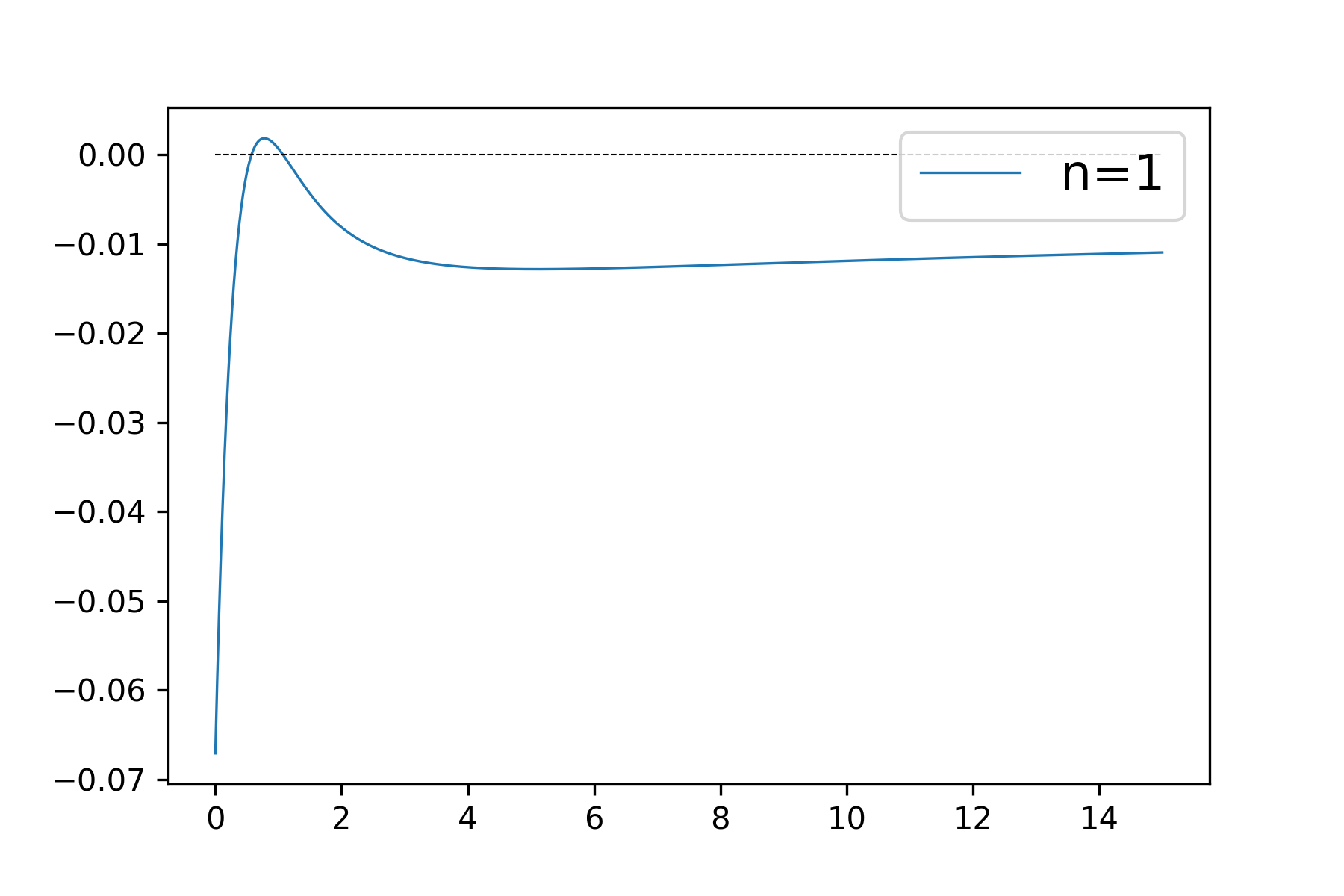}
\caption{Having unstable area}\label{eigs1}
\end{subfigure}
\caption{Eigenvalues as a function of $\tau_M$ for $n=1$}\label{eigs}
\end{figure}

Figure \ref{eigs} plots the eigenvalue \eqref{ev} on the vertical axis for the frequency $n=1$ with $\tau^{M}$ on the horizontal axis. Both cases share the same parameter values $\mu=0.5$, $\eta=2.0$, $\sigma=3.0$, but $\tau^A=2.5$ in Figure \ref{eigs}\subref{eigs0} and $\tau^A=2.0$ in Figure \ref{eigs}\subref{eigs1}. In Figure \ref{eigs}\subref{eigs0}, the eigenvalue is negative for any value of $\tau^{M}$, i.e., the mode with $n=1$ is is stable over all values of $\tau^M$. On the other hand, Figure \ref{eigs}\subref{eigs1} shows a more general situation. In this case, the eigenvalue is negative at sufficiently high values of $\tau^M$, positive at medium values of $\tau^{M}$, and negative again at low values of $\tau^M$. As a result, the eigenvalue \eqref{ev} has two critical points $\ul{\tau}^M < \ol{\tau}^{M}$ at which the eigenvalue \eqref{ev} equals to zero. There, the mode with $n=1$ becomes unstable under values of $\tau^M\in(\ul{\tau}^M, \ol{\tau}^M)$.

\begin{figure}[H]
\centering
\includegraphics[width=0.7\columnwidth]{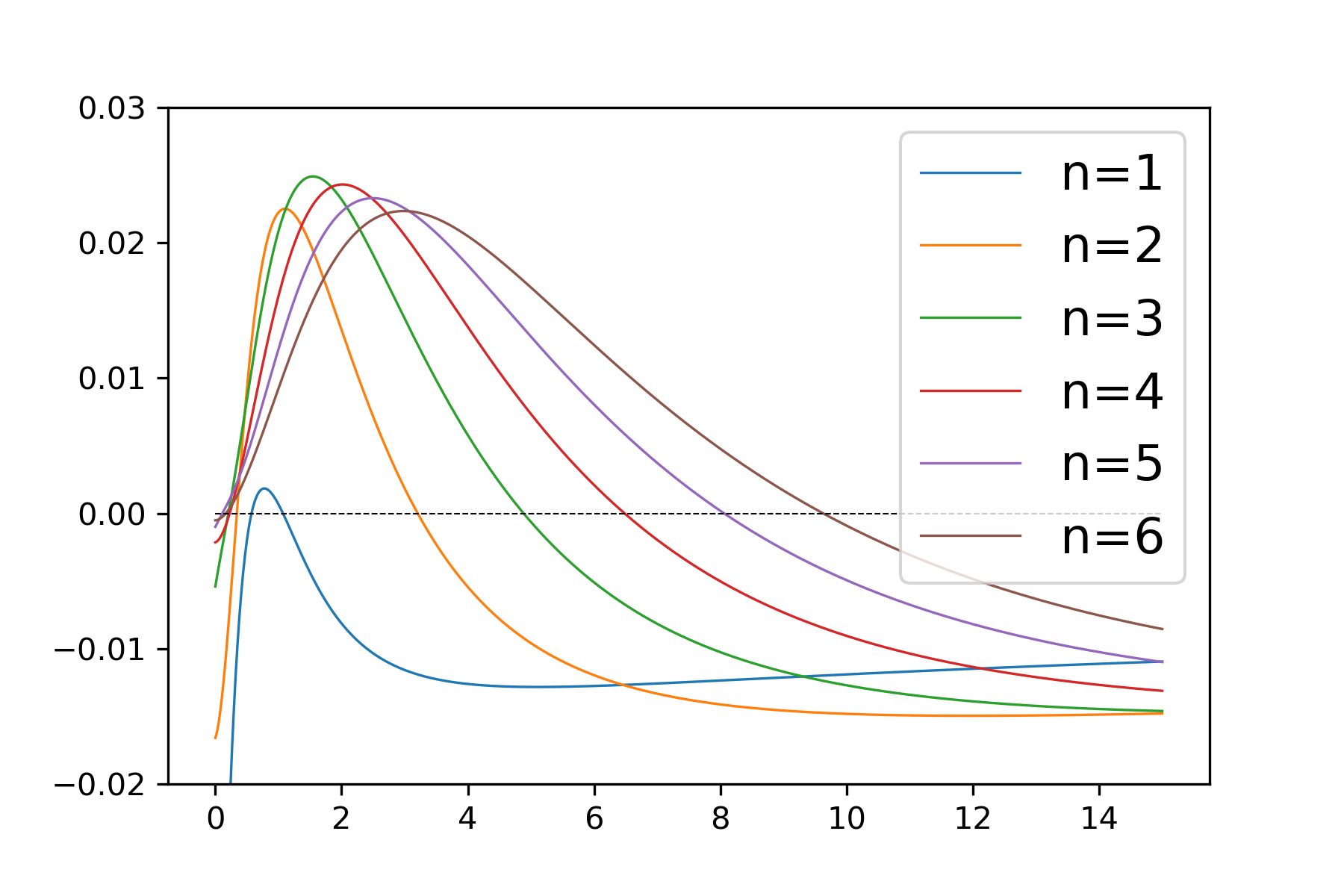}
\caption{Eigenvalues as functions of $\tau_M$ for $n=1$ to $6$}\label{eigsmore}
\end{figure}

Figure \ref{eigsmore} shows the eigenvalues for more frequencies. As the absolute value of frequencies $|n|$ increases, the upper critical point $\ol{\tau}^M$ is observed to move to the right. This means that as manufacturing transport costs decrease, modes with larger $|n|$ destabilize before those with smaller $|n|$. The similar property would be observed for various CP models on a circle which do not consider differentiated agricultural goods.\footnote{See, for example, \citet{TabuThis}, \citet{AkaTakaIke}, \citet{IkeOnTaka}, \citet{OhtakeYagi_point}, and \citet{Ohtake2023cont}.} Contrary to the obviousness of the numerical result, it is difficult to give a complete analytical proof of the rightward shift of $\ol{\tau}^M$ as $|n|$ increases.\footnote{For a rigorous proof, it is necessary to compute analytically how solutions of a quadratic equation \eqref{QH} behaves with respect to change of  $n$. If $\alpha\to 0$, then the coefficients of the quadratic equation do not depend on $H^\alpha_n$ and thus the solutions do not depend on $n$. In this case, it is easy to prove the rightward shift of $\ol{\tau}^M$ as Theorem \ref{th:monoexpalpha=0} with the change from $|n|$ to $|n|+2$. When $\alpha\neq 0$, however, the solutions of the quadratic equation depend on $H^\alpha_n$ in a complicated manner, as expressed in \eqref{b}, \eqref{B}, and \eqref{omhat}. As a result, it is still difficult to prove analytically the rightward shift when $|n|$ changes to $|n|+1$.
} For $\alpha=0$, however, we can obtain the following theorem which states that the $\ol{\tau}^M$ moves to the right as the absolute value of the frequency changes from $|n|$ to $|n|+2$. See Subsection \ref{proof:th:monoexpalpha=0} for the proof.
\begin{theo}\label{th:monoexpalpha=0}
If $\alpha=0$, then the lower critcal point $\ul{\tau}^M$ equals to zero, and the upper critical point $\ol{\tau}^M$ increases as the absolute value of frequencies changes from $|n|$ to $|n|+2$.
\end{theo}

The homogeneous stationary solution becomes unstable when decreasing $\tau^M$ pass the upper critical point $\ol{\tau}^M$, but the homogeneous stationary solution becomes stable again when further decreasing $\tau^M$ pass the lower critical point $\ul{\tau}^M$. This ``redispersion'' is consistent with the result of the two-region model by \citet[Chapter 7]{FujiKrugVenab}. 

\subsection{Critical curves}\label{subsec:ccurves}
\begin{figure}[H]
\begin{subfigure}[H]{0.5\columnwidth}
\includegraphics[width=\columnwidth]{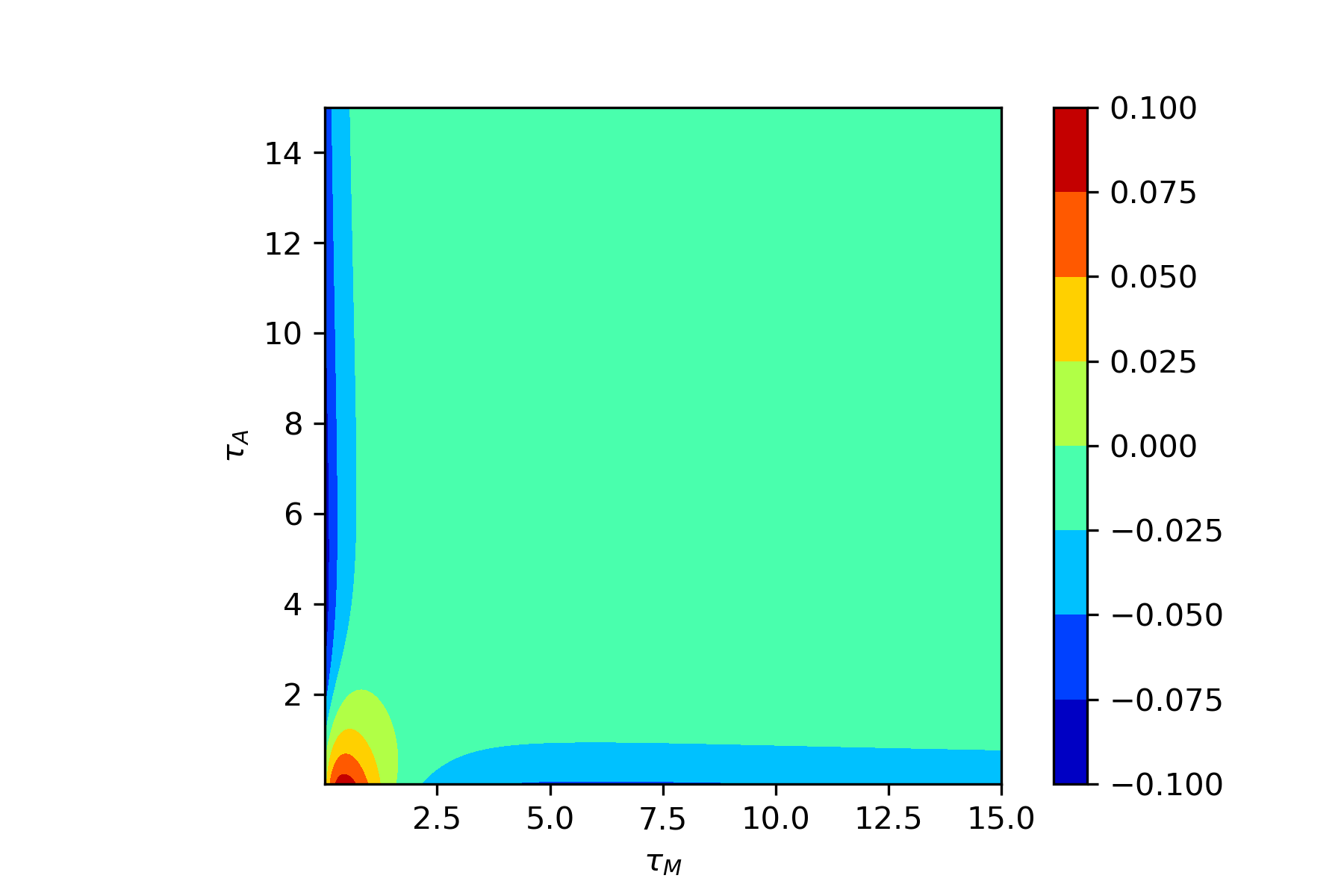}
\caption{$n=1$}
\end{subfigure}
\begin{subfigure}[H]{0.5\columnwidth}
\includegraphics[width=\columnwidth]{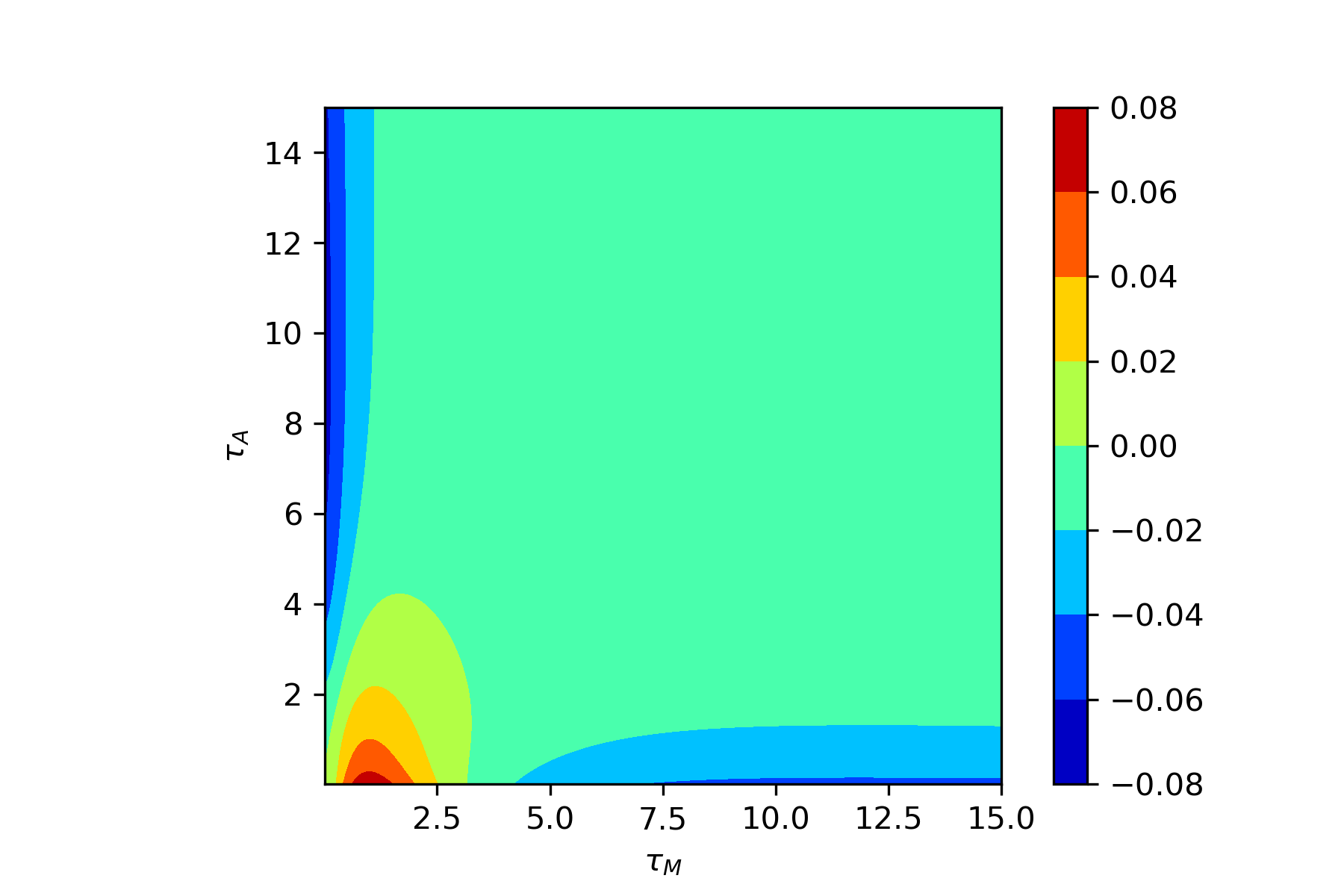}
\caption{$n=2$}
\end{subfigure}
\begin{subfigure}[H]{0.5\columnwidth}
\includegraphics[width=\columnwidth]{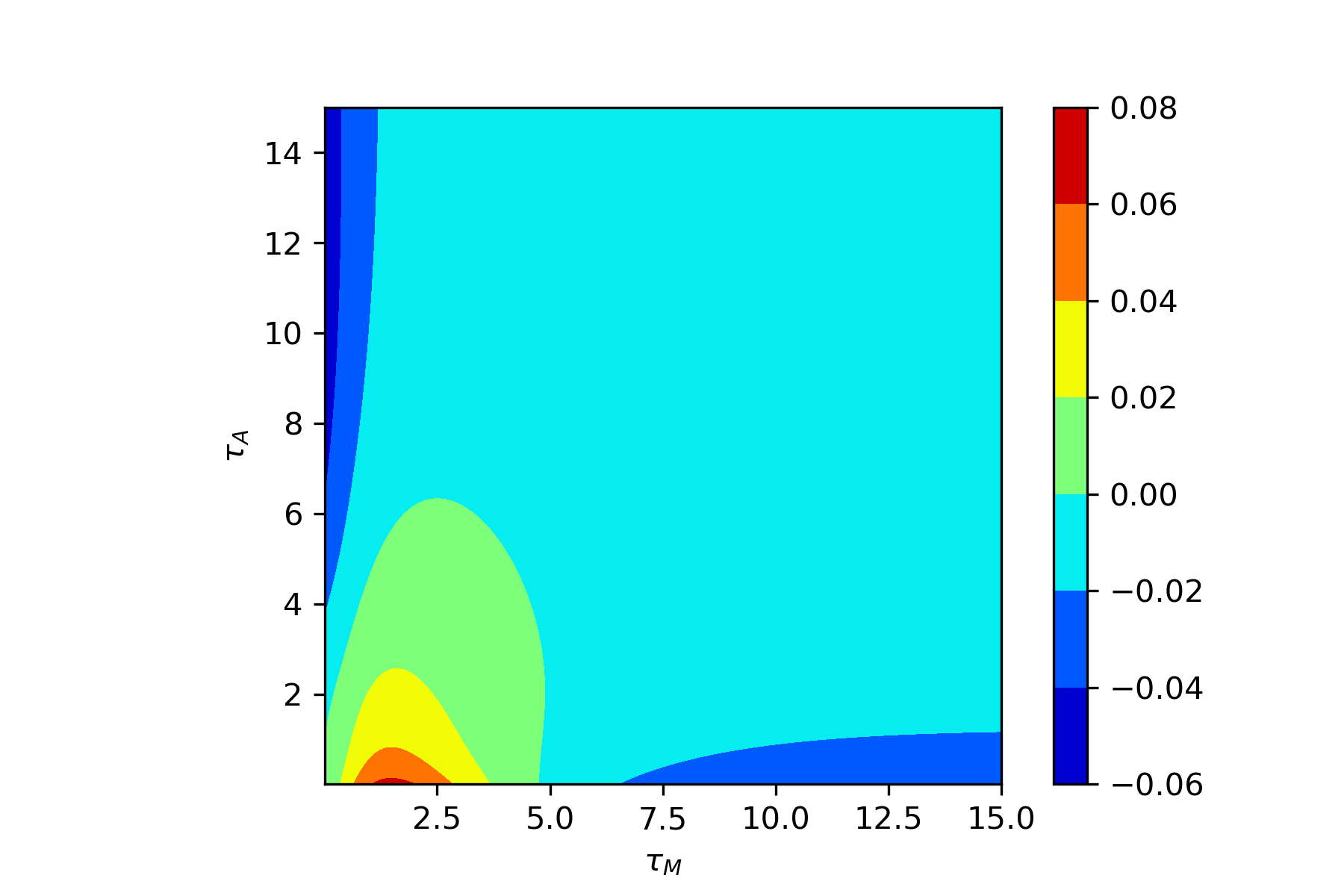}
\caption{$n=3$}
\end{subfigure}
\begin{subfigure}[H]{0.5\columnwidth}
\includegraphics[width=\columnwidth]{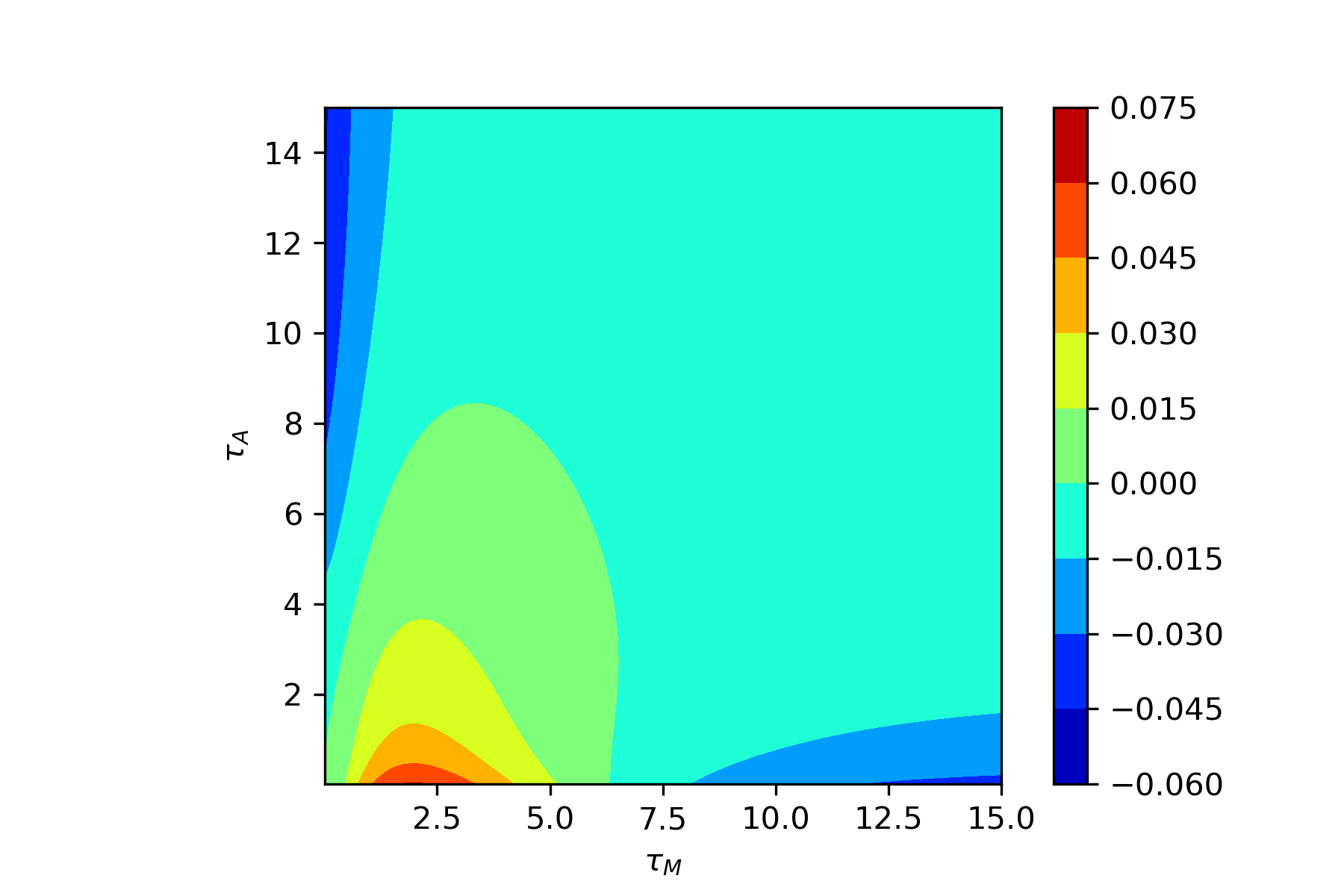}
\caption{$n=4$}
\end{subfigure}
\begin{subfigure}[H]{0.5\columnwidth}
\includegraphics[width=\columnwidth]{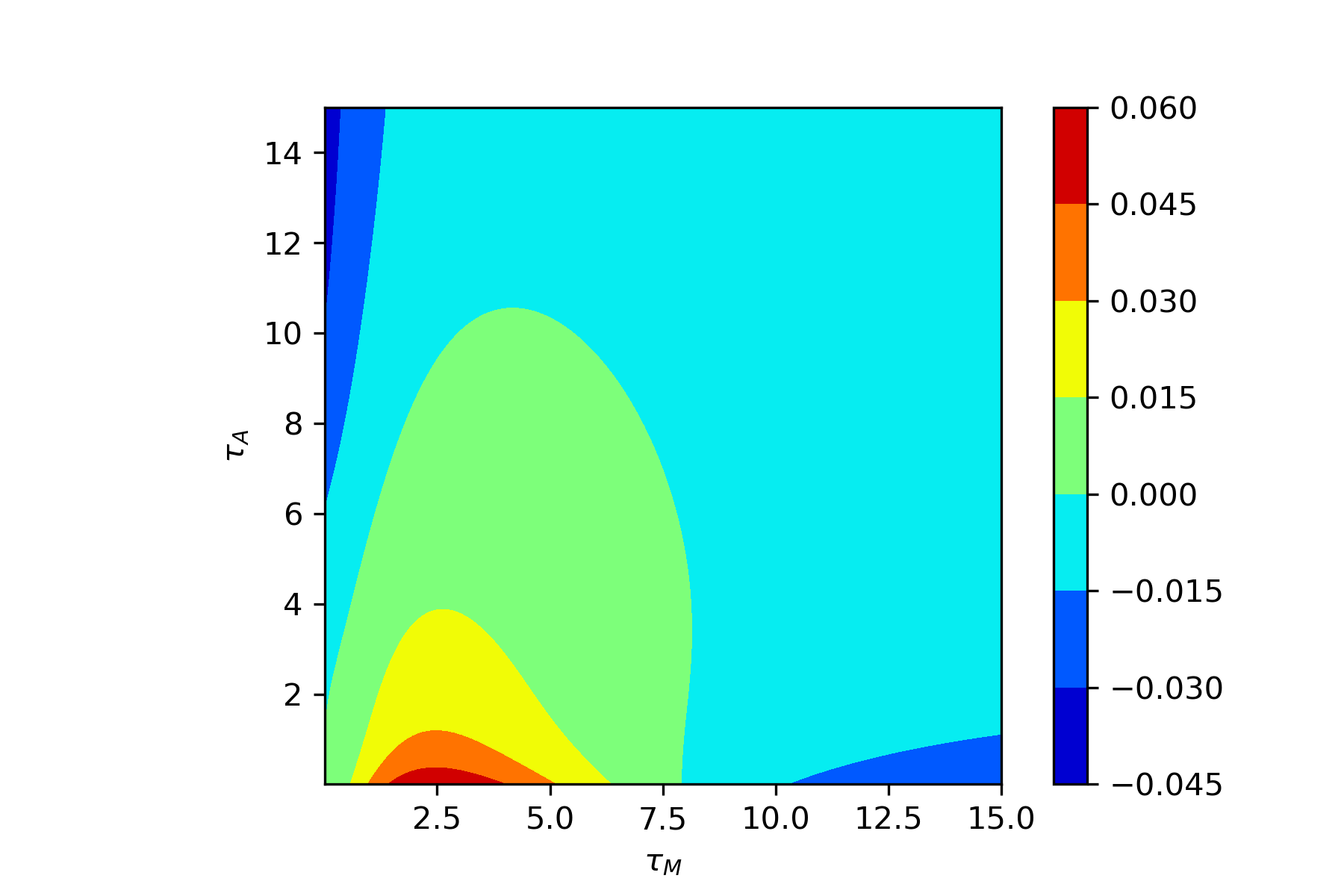}
\caption{$n=5$}
\end{subfigure}
\begin{subfigure}[H]{0.5\columnwidth}
\includegraphics[width=\columnwidth]{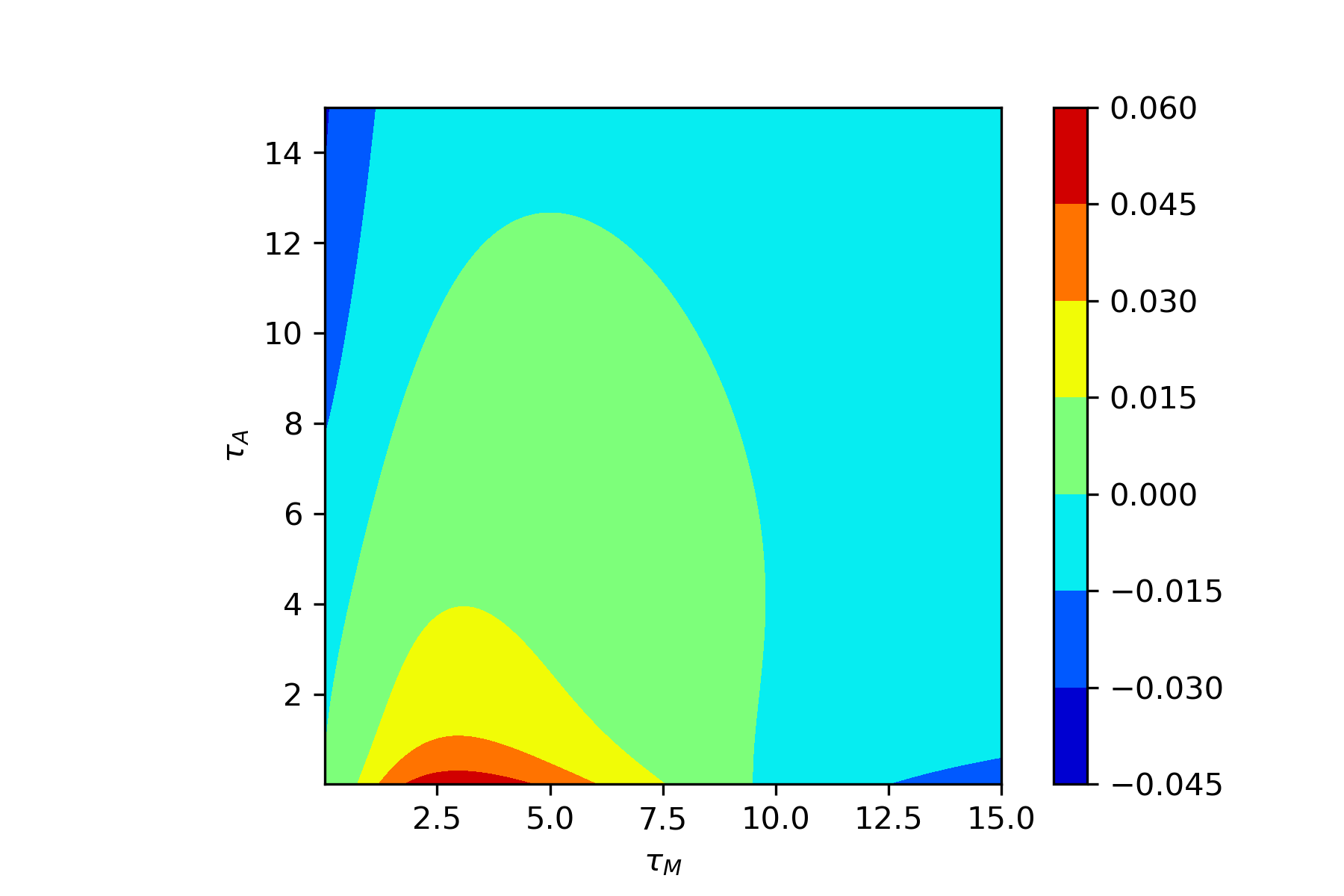}
\caption{$n=6$}
\end{subfigure}
\caption{Heatmaps of the eignevalue in $(\tau_M, \tau_A)$-space}\label{figs:hm_tauA}
\end{figure}

\begin{figure}[H]
\centering
\includegraphics[width=0.7\columnwidth]{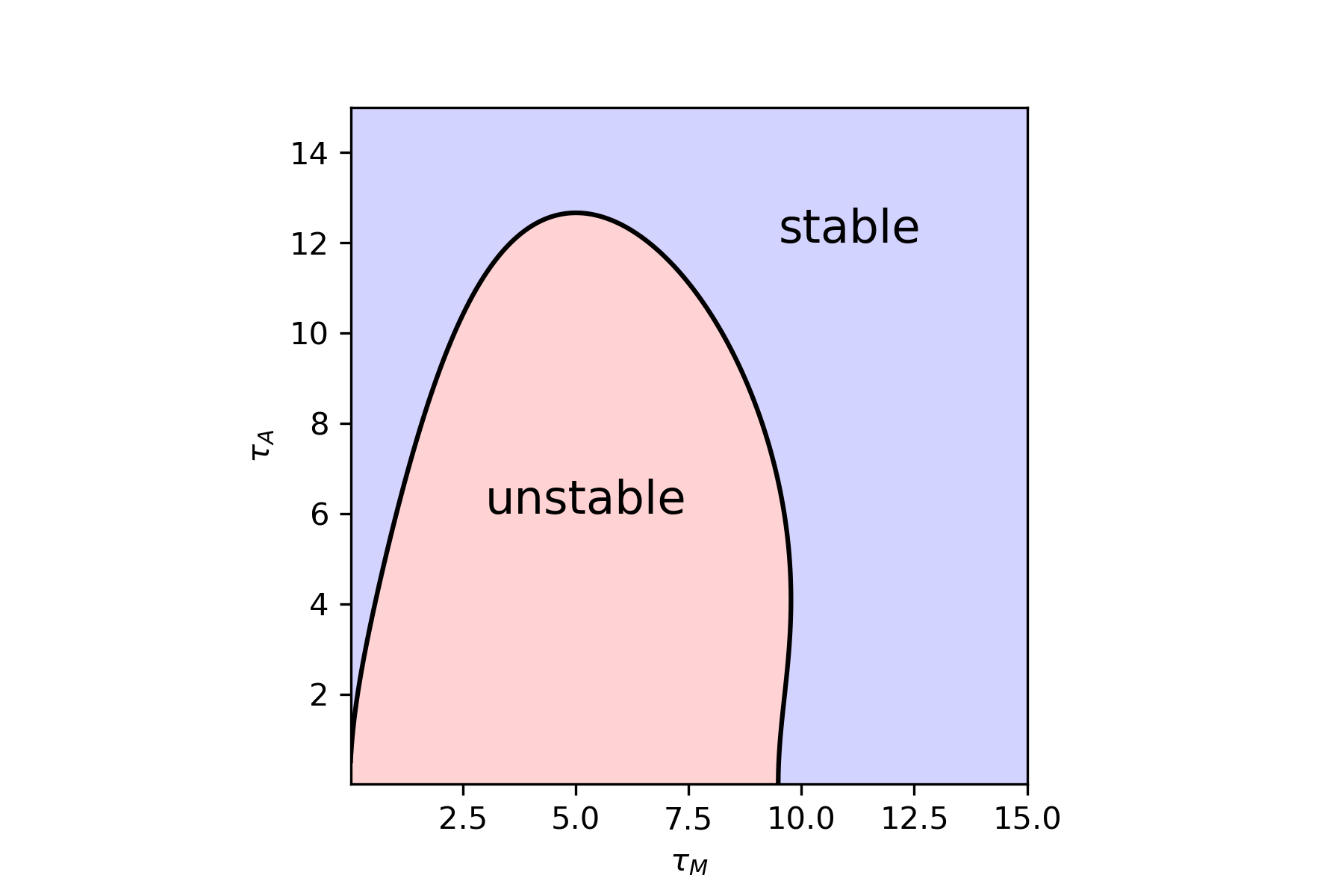}
\caption{Critical curve and two domains}\label{fig:cv_u_s_domains}
\end{figure}

\begin{figure}[H]
\centering
\includegraphics[width=0.7\columnwidth]{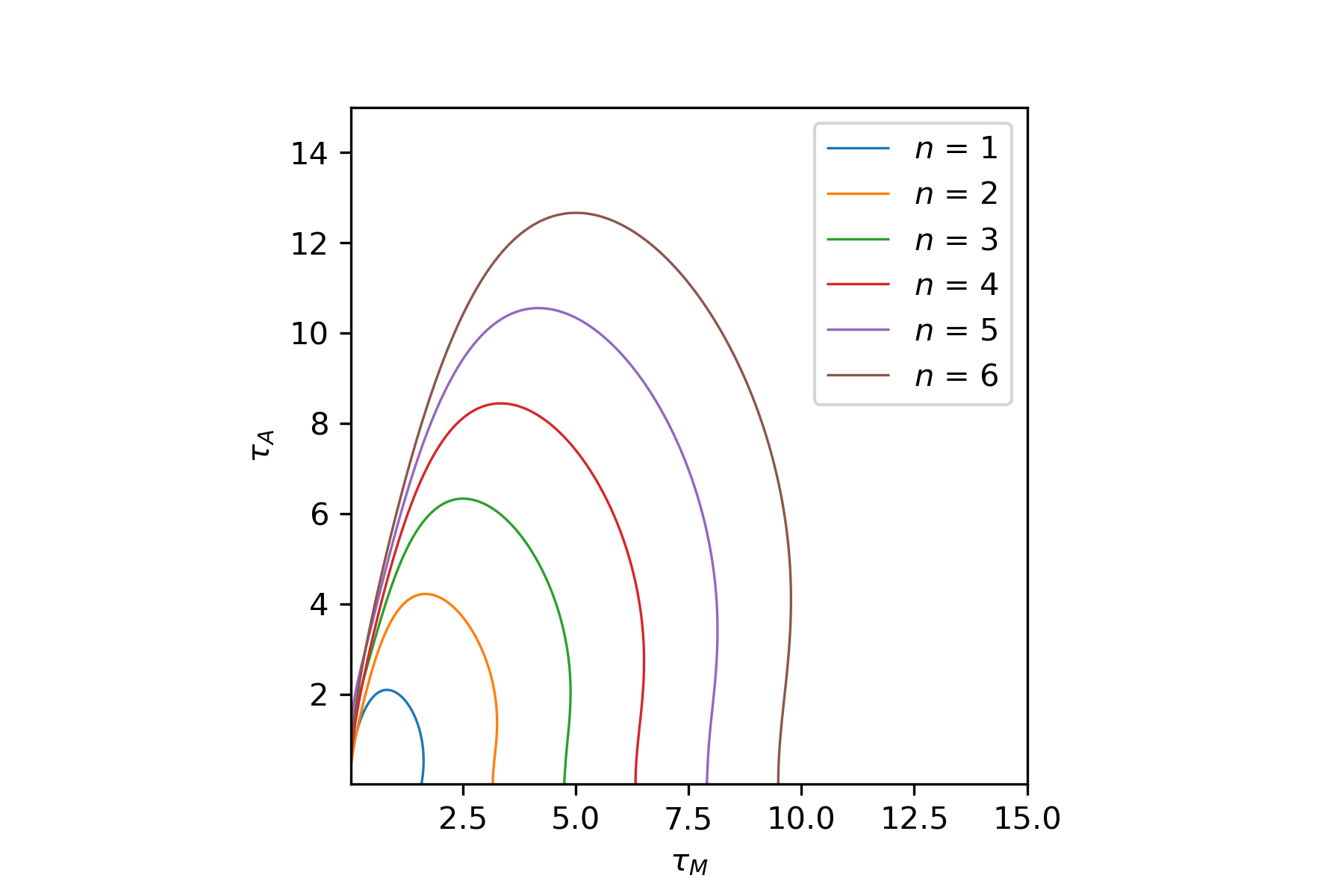}
\caption{Critical curves for $n=1$ to $6$}\label{fig:ta_ccs}
\end{figure}

Let us examine how the eigenvalue relates to the agricultural transport cost. Figure \ref{figs:hm_tauA} shows heatmaps of the eigenvalue in $(\tau^M, \tau^A)$-plane under $\mu=0.5$, $\eta=2.0$, and $\sigma=3.0$.\footnote{All the figures in this subsection are drawn by Matplotlib ver. 3.6.2.} For each $n=1$ to $6$, there exists a curve that divides the plane into two domains, one for positive eigenvalues and another one for negative eigenvalues. Let us call this curve the {\it critical curve} and the two domains the {\it unstable} and {\it stable domains}, respectively (See Figure \ref{fig:cv_u_s_domains}). In Figure \ref{fig:ta_ccs}, the critical curves are drawn for each $n$. It appears that the unstable domain expands monotonically as $n$ increases. However, as a matter of fact, on the left side of the figure, i.e., in the range where $\tau^M$ is relatively small, ciritical curves for different frequencies intersect slightly in some cases, thus the expansion is not necessarily ``monotonic".\footnote{For example, in Figure \ref{fig:ta_ccs}, the critical curves for $n=1$ and $n=2$ intersect slightly. From numerical observations, one would expect that the critical curves for $n$ and $n+2$ would not intersect, but this has not been proven.} Even so, on the right side of the figure, where $\tau^M$ is relatively large, monotonous expansions of the unstable domain can clearly be seen. In fact, the critical curve intersections are so trivial that it might be reasonable to say that the unstable domain expands ``in general'' monotonically for $n$.

Hence, we see that, in general, for each value of $\tau^M$, sufficiently decreasing values of $\tau^A$ destabilize modes having smaller frequencies, thus promoting agglomeration. That lower agricultural transport costs promote agglomeration is consistent with the results of \citet[pp.107-111]{FujiKrugVenab}. This can be understood to be because sufficiently low agricutural transport costs facilitate the import of agricultural goods from other regions, thus recovering the difficulties in transporting agricultural goods caused by agglomeration.\footnote{\citet[p.140]{ZenTaka} give a similar interpretation for a qusi-linear two-region model proposed in \citet{PicaZeng05}.}

\begin{figure}[H]
\begin{subfigure}[H]{0.5\columnwidth}
\includegraphics[width=\columnwidth]{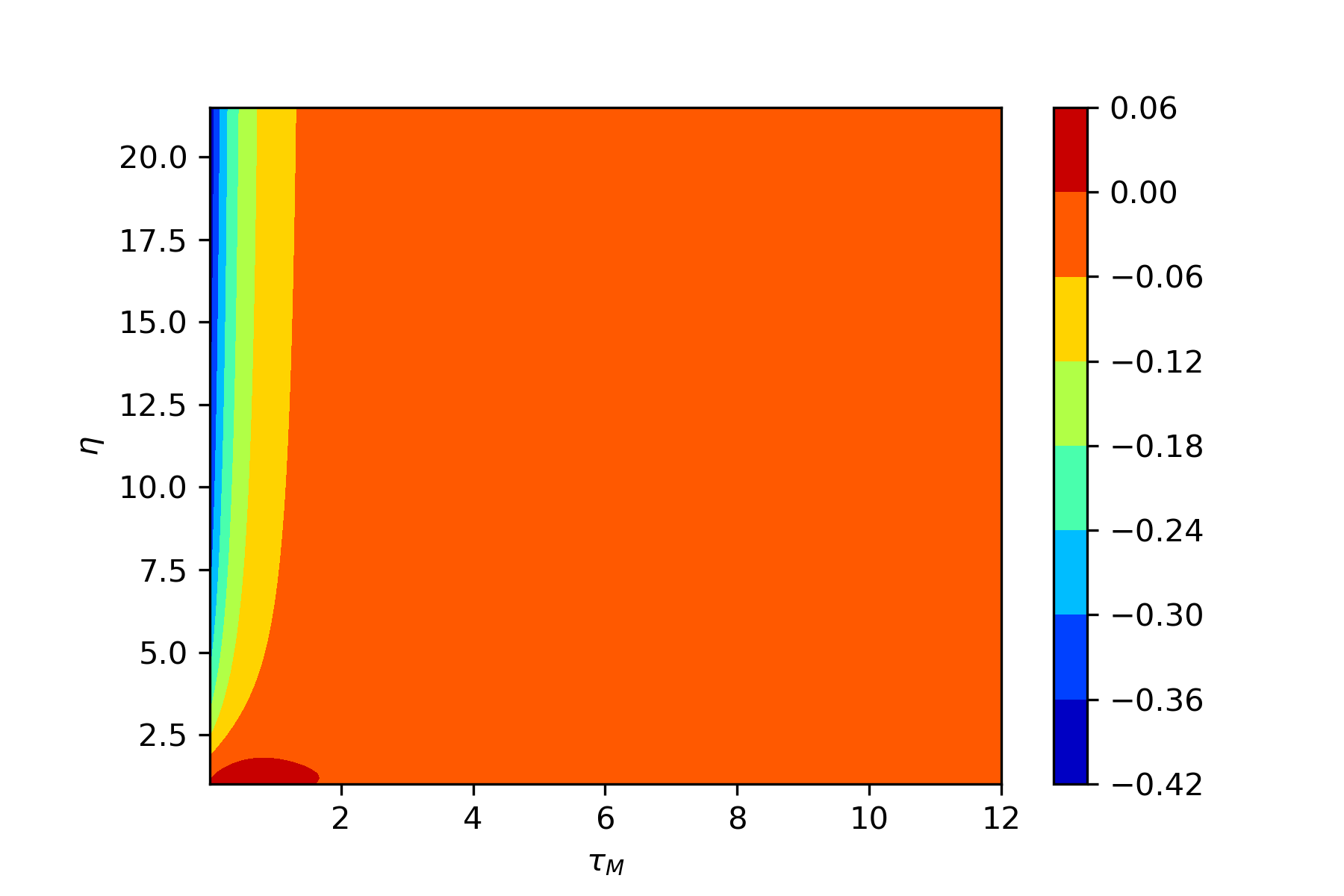}
\caption{$n=1$}
\end{subfigure}
\begin{subfigure}[H]{0.5\columnwidth}
\includegraphics[width=\columnwidth]{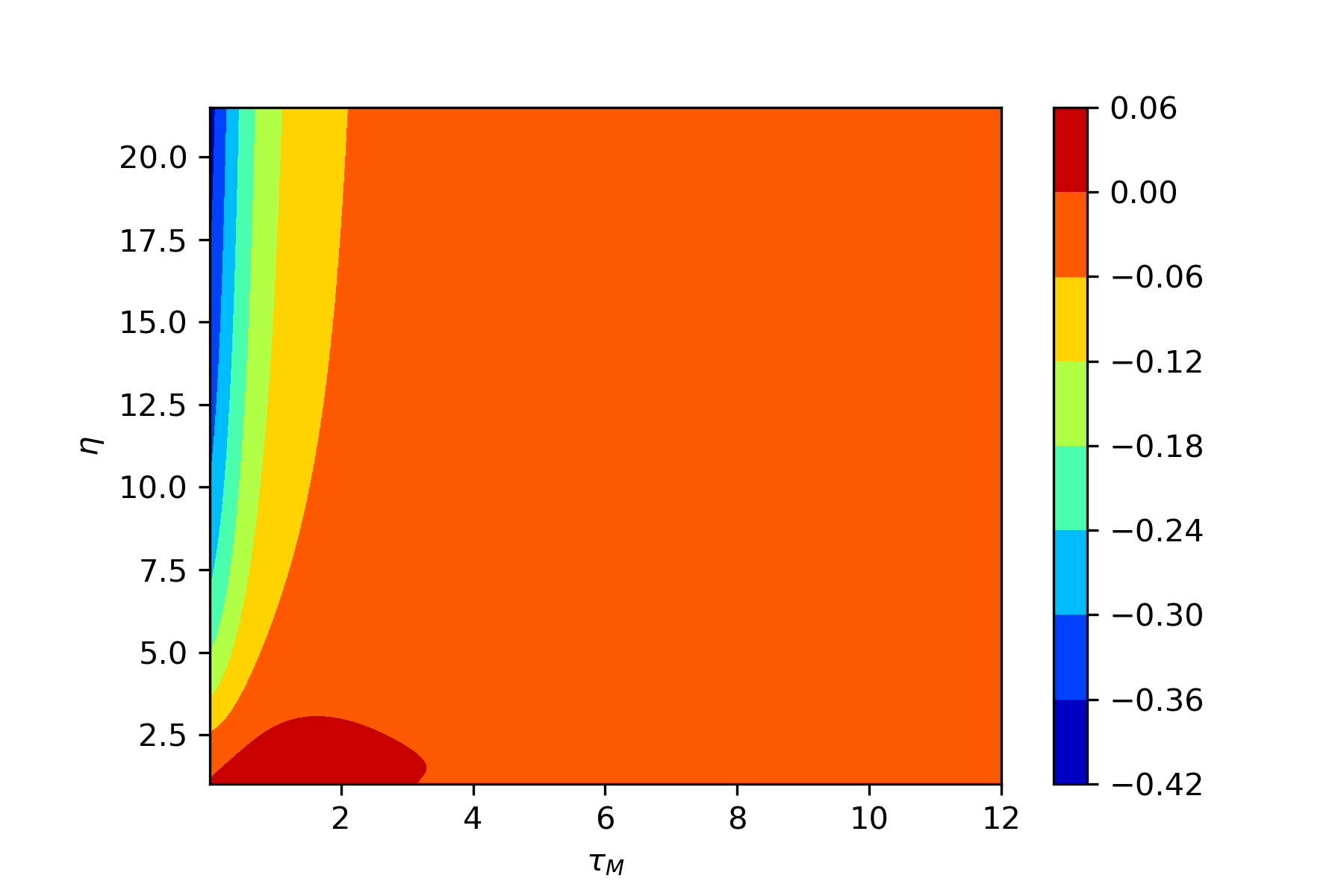}
\caption{$n=2$}
\end{subfigure}
\begin{subfigure}[H]{0.5\columnwidth}
\includegraphics[width=\columnwidth]{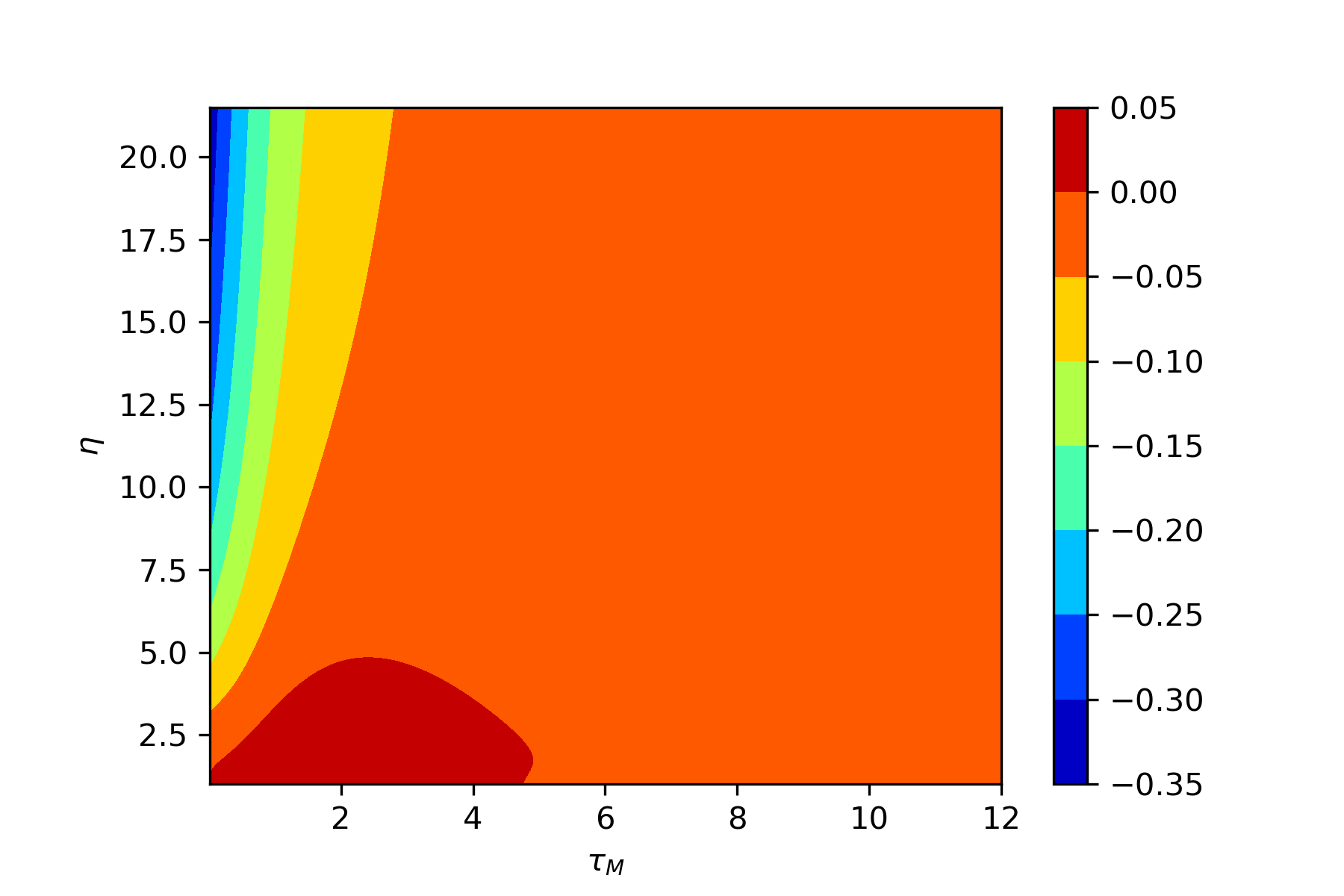}
\caption{$n=3$}
\end{subfigure}
\begin{subfigure}[H]{0.5\columnwidth}
\includegraphics[width=\columnwidth]{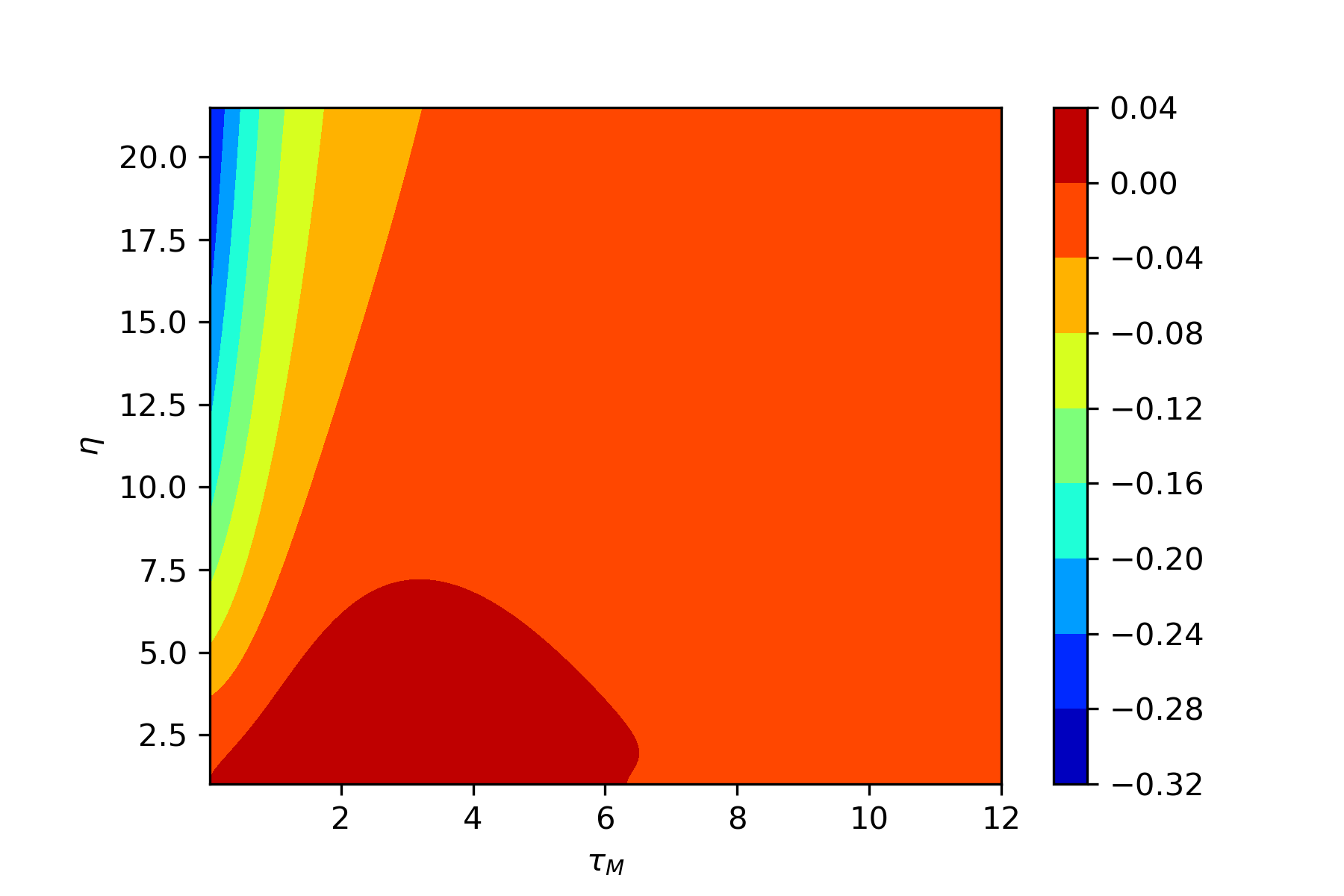}
\caption{$n=4$}
\end{subfigure}
\begin{subfigure}[H]{0.5\columnwidth}
\includegraphics[width=\columnwidth]{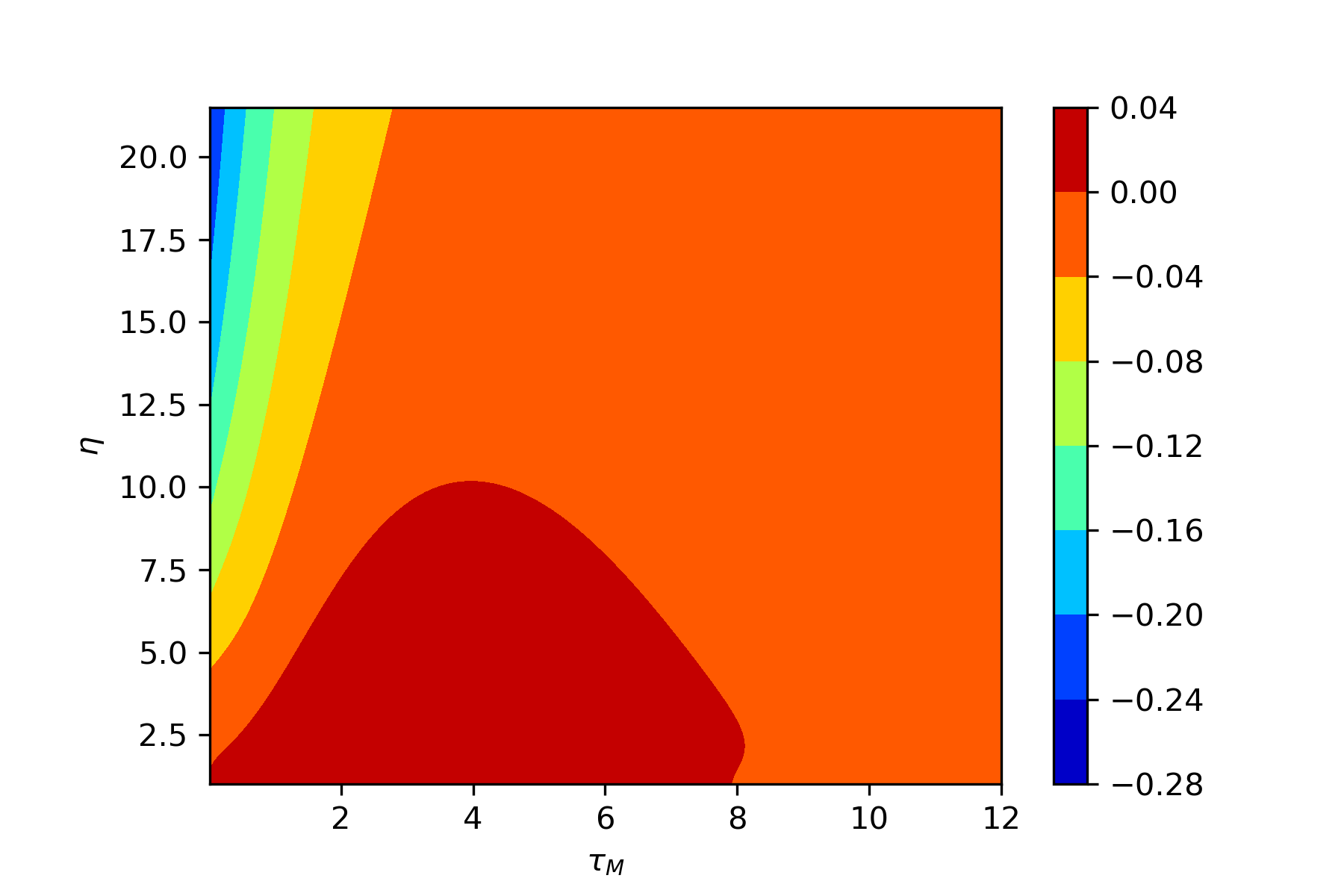}
\caption{$n=5$}
\end{subfigure}
\begin{subfigure}[H]{0.5\columnwidth}
\includegraphics[width=\columnwidth]{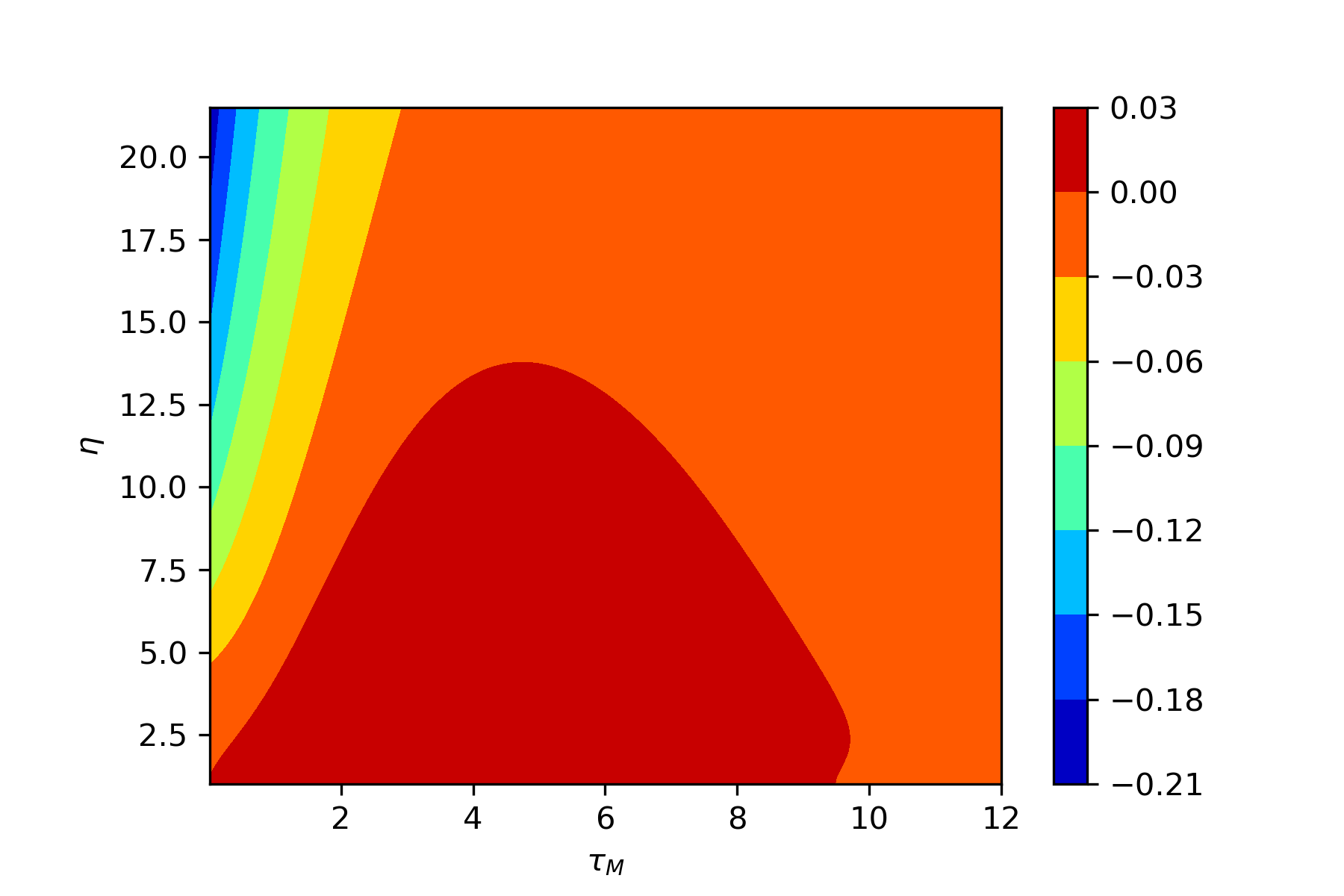}
\caption{$n=6$}
\end{subfigure}
\caption{Heatmaps of the eigenvalue in $(\tau_M, \eta)$-space}\label{figs:hm_eta}
\end{figure}

\begin{figure}[H]
\centering
\includegraphics[width=0.7\columnwidth]{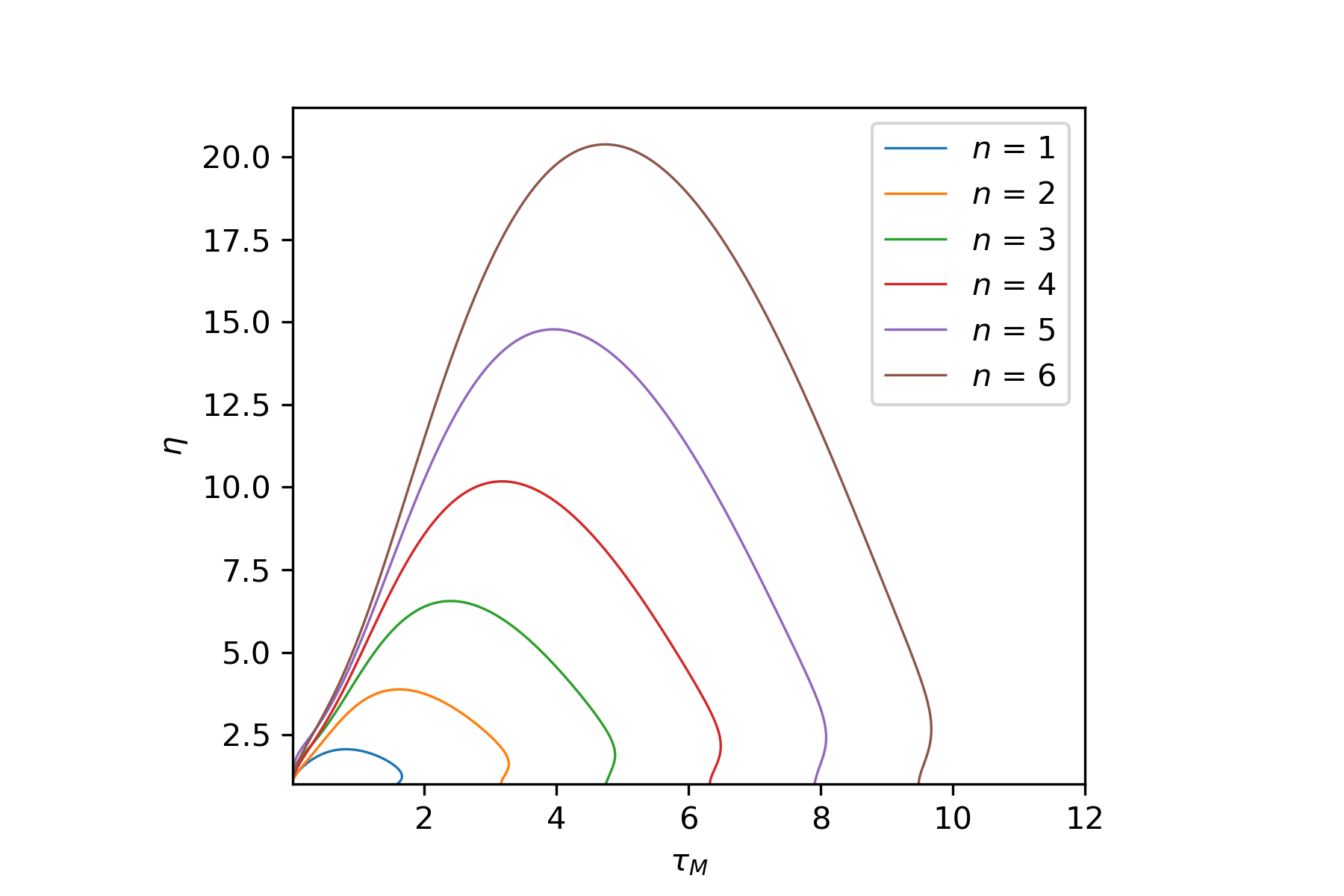}
\caption{Critical curves for $n=1$ to $6$}\label{fig:eta_ccs}
\end{figure}

Next, we examine how the eigenvalue relates to the preference for agricultural variety. For $\eta>1$, the heatmaps of the eigenvalue and the critical curves for the frequencies $n=1$ to $6$ under $\mu=0.5$, $\sigma=3.0$, and $\tau_A=2.0$ are shown in Figure \ref{figs:hm_eta} and Figure \ref{fig:eta_ccs}, respectively. As in the case of the agricultural transport costs, the critical curves for different $n$ are sometimes slightly intersecting.\footnote{We also see that, in Figure \ref{fig:eta_ccs}, the critical curves for $n=1$ and $n=2$ intersect slightly for example. From numerical observations, one would again expect that the critical curves for $n$ and $n+2$ would not intersect, but this has not been proven.} Again, however, it would be safe to say that the unstable domain expands in general monotonically as $n$ increases.

From Figure \ref{fig:eta_ccs}, it can be seen that for each value of $\tau^M$, sufficiently decreasing values of $\eta$ destabilize eigenfunctions having smaller $n$, and therefore, early stage agglomerations in a smaller number of regions are expected to be observed. This means that the greater the preference for agricultural variety, the more agglomeration is promoted. This can be intuitively interpreted as the decline in the price indices of the agricultural goods due to the stronger preference for agricultural variety\footnote{Actually, it is easy from \eqref{piA} to see that $\frac{d}{d\eta}G^A>0$.} offsetting an increase in the prices of the agricultural goods in the agglomerated regions, allowing the workers to enjoy the benefits of agglomeration.\footnote{\citet[p.141]{ZenTaka} give a similar interpretation for a quasi-linear two-region model proposed in \citet{PicaZeng05}.}

\section{Numerical examples of asymptotic behavior}\label{sec:num}
In this section, we numerically explore the so-called asymptotic behavior of solutions of \eqref{consys}, that is, what distribution the solutions asymptotically approach after a long period of time.\footnote{All numerical simulations in this section were performed by using Julia (\citet{be2017julia}) ver. 1.10.2.}

\subsection{Numerical settings}
We discretize the time and the space variables in the following manner. By a certain width $\varDelta t>0$, the time is discretized into nodes $t_k=k\varDelta t$ for $k=0, 1, 2, \cdots$. The interval $[-\pi, \pi]$ is discretized into $I$ nodes $\theta_i = -\pi+(i-1)\varDelta\theta$, where $\varDelta\theta=2\pi/I$ for $i=1,2,\cdots,I$. In the following simulation, we set $\varDelta t=0.01$ and $I=128$. Any function $f$ on $[0,\infty)\times S$ is approximately computed only for values on these nodes. Let $f^k_i=f(t_k, x_i)$ and $f^k=[f_1^k,f_2^k,\cdots,f_I^k]$.

The numerical scheme we use is explicit with respect to time evolution. That is, we only need an approximated population distribution $\lambda^k$ at time step $k$ to obtain $\lambda^{k+1}$ (and its associated approximations to the other unknown functions) at time step $k+1$. Starting from an initial distribution $\lambda^0$, time evolution is simulated one after another, and finally when $\left\|\lambda^k-\lambda^{k+1}\right\|_\infty<10^{-10}$, it is regarded as reaching a stationary solution and the computation is stopped.\footnote{$\|f\|_\infty:=\max \left\{|f_1|,|f_2|,\cdots,|f_I|\right\}$.} The approximate stationary solution obtained in this way is denoted by $\lambda^*$. See Subsection \ref{subsec:numsch} for details of the numerical scheme. The source code for the simulation is available at \url{https://github.com/k-ohtake/contspace-cpmodel-agriculture}.

\subsection{Spiky agglomeration and redispersion}
\begin{figure}[H]
\centering
\includegraphics[width=0.6\columnwidth]{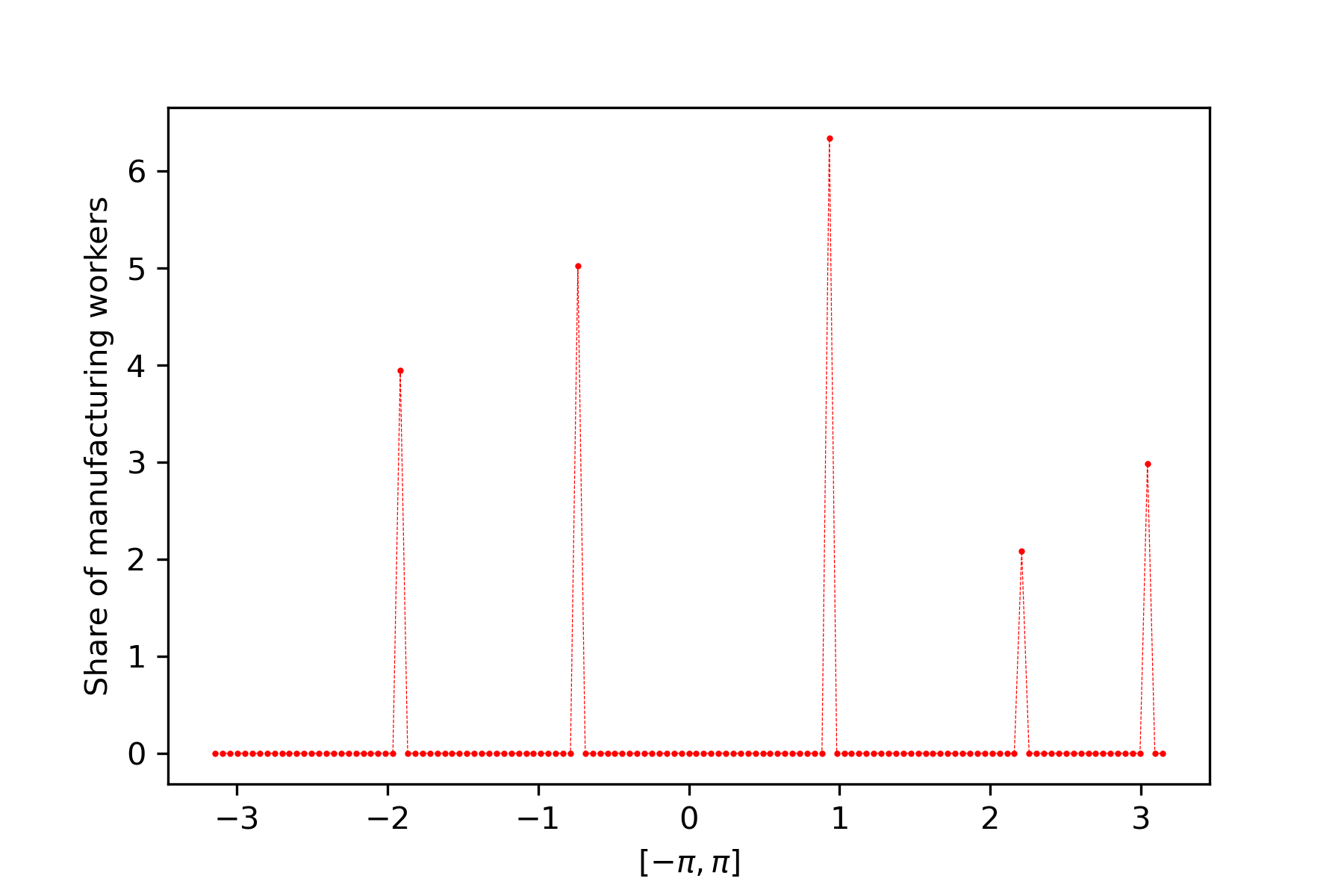}
\caption{Stationary solution with five spikes}\label{fig:stsol1}
\end{figure}

\begin{figure}[H]
\centering
\includegraphics[width=0.6\columnwidth]{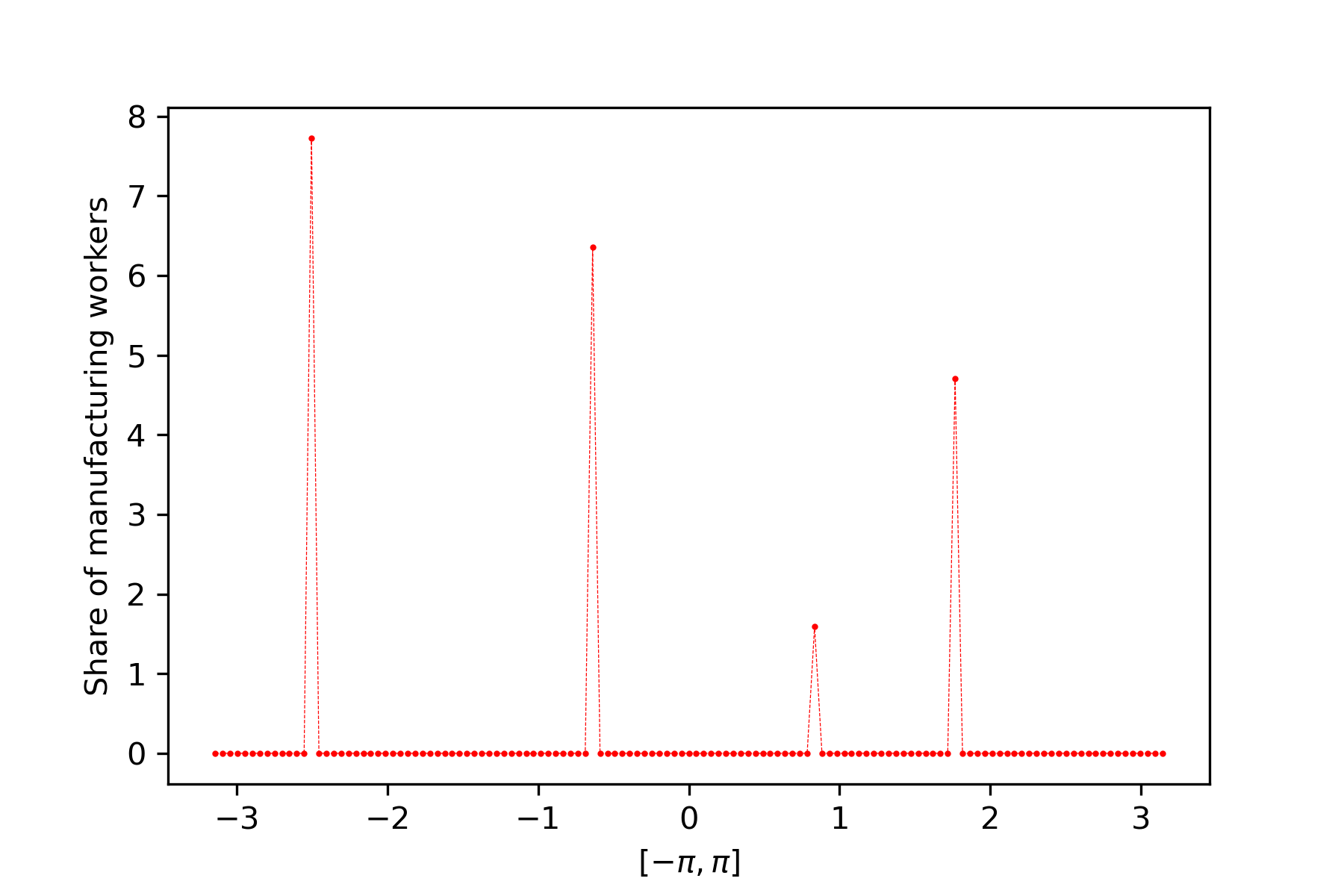}
\caption{Stationary solution with four spikes}\label{fig:stsol2}
\end{figure}

\begin{figure}[H]
\centering
\includegraphics[width=0.6\columnwidth]{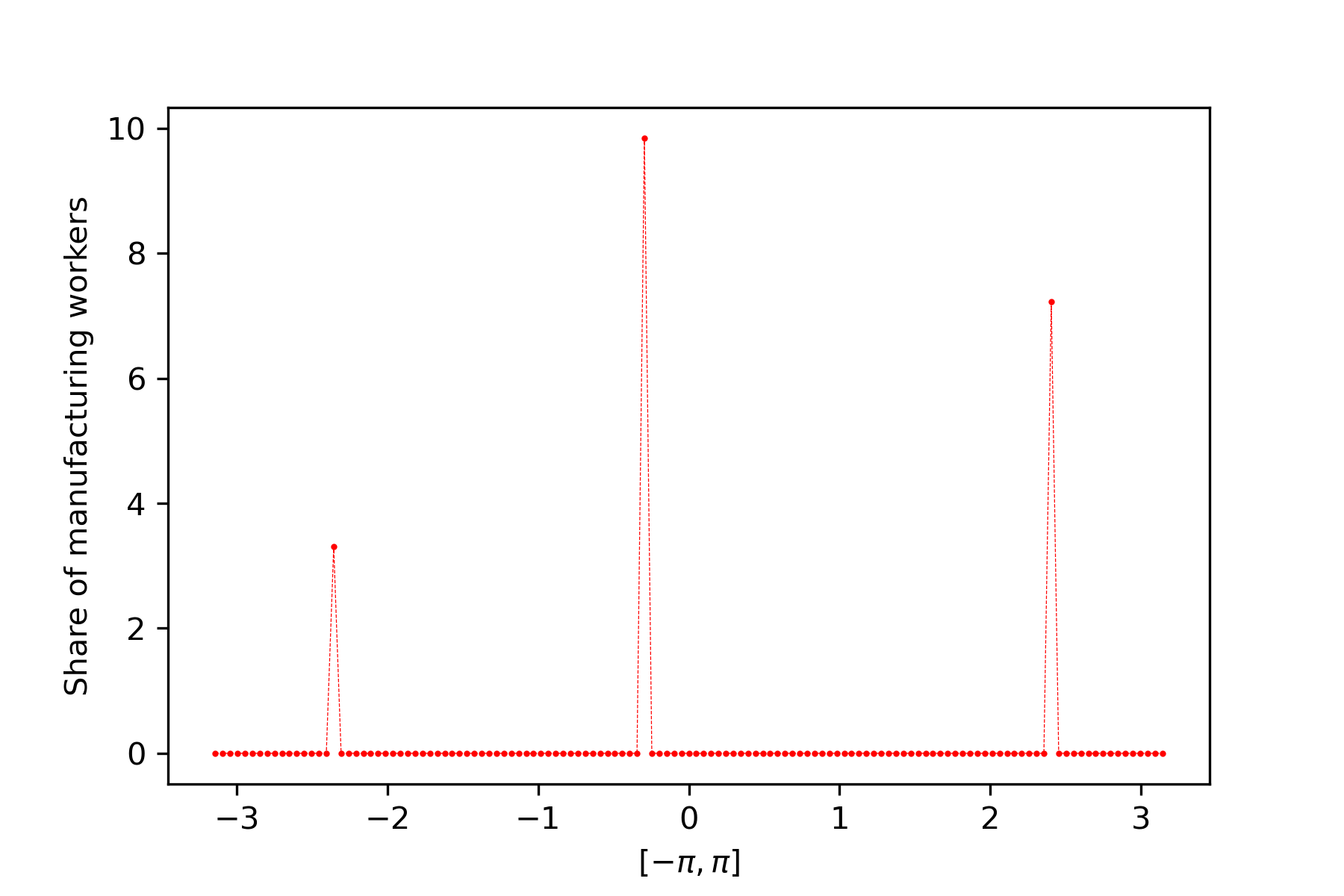}
\caption{Stationary solution with three spikes}\label{fig:stsol3}
\end{figure}

For initial population distributions generated by adding small random purturbations to the homogeneous distribution $\ol{\lambda}=1/(2\pi\rho)$, Figures \ref{fig:stsol1} to \ref{fig:stsol3} show $\lambda^*$ of non-homogeneous stationary solutions on $[-\pi, \pi]$.\footnote{The parameters are set as follows.\\
Fig. \ref{fig:stsol1}: $\rho=1.0$, $\mu=0.5$, $\sigma=3.0$, $\eta=2.0$, $\tau^A=2.0$, and $\tau^M=5.0$.\\
Fig. \ref{fig:stsol2}: $\rho=1.0$, $\mu=0.5$, $\sigma=3.0$, $\eta=2.0$, $\tau^A=2.0$, and $\tau^M=4.0$.\\
Fig. \ref{fig:stsol3}: $\rho=1.0$, $\mu=0.5$, $\sigma=3.0$, $\eta=2.0$, $\tau^A=2.0$, and $\tau^M=2.5$.\\
In these figures, the actual computed values are indicated by the dots. The dashed lines are interpolation for the plot.} Evidently, it can be seen that $\lambda^*$ has multiple spikes.


\begin{figure}[H]
 \begin{subfigure}{0.5\columnwidth}
  \centering
  \includegraphics[width=\columnwidth]{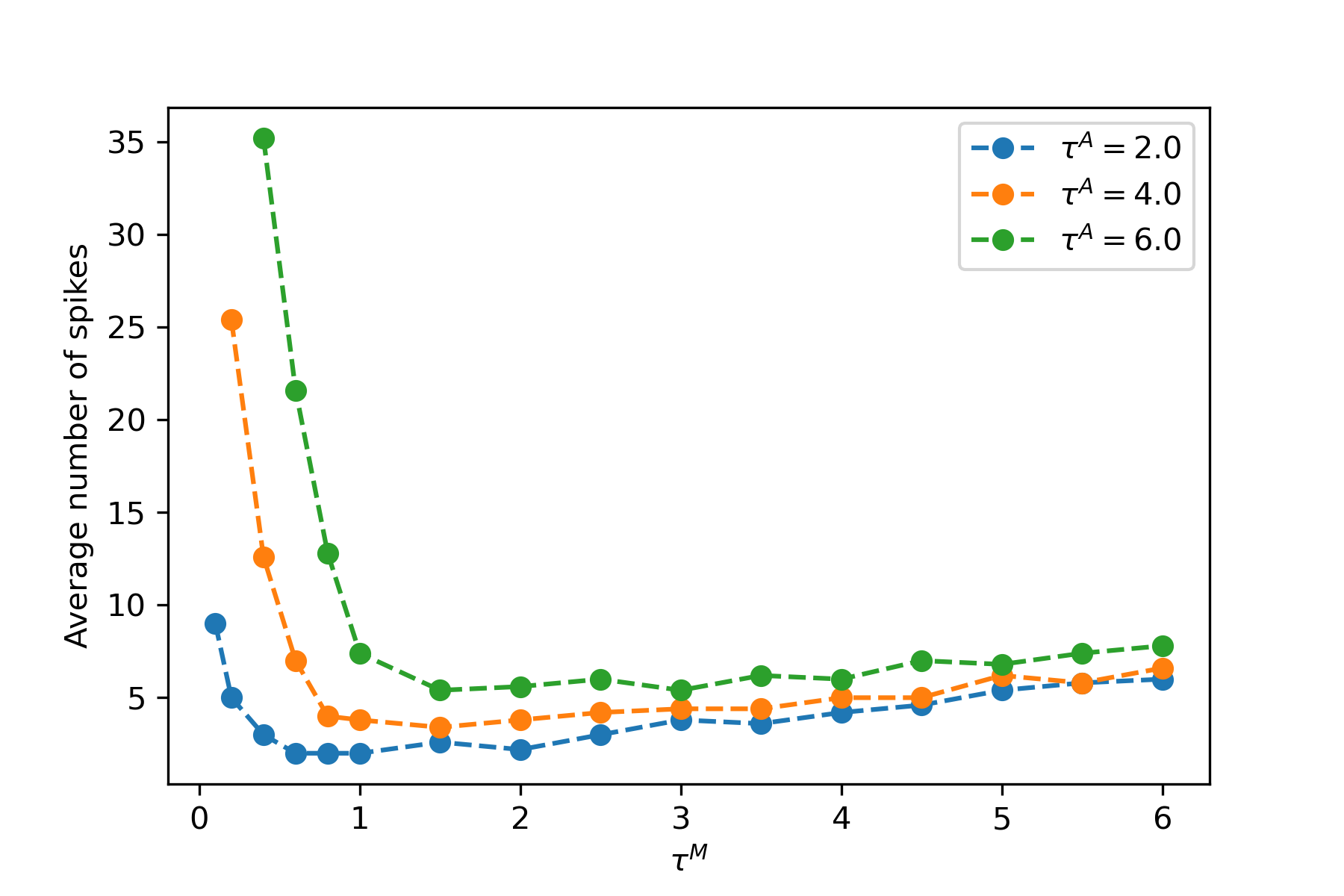}
  \caption{Under different values of $\tau^A$}\label{subfig:avenumspta}
 \end{subfigure}
 \begin{subfigure}{0.5\columnwidth}
  \centering
  \includegraphics[width=\columnwidth]{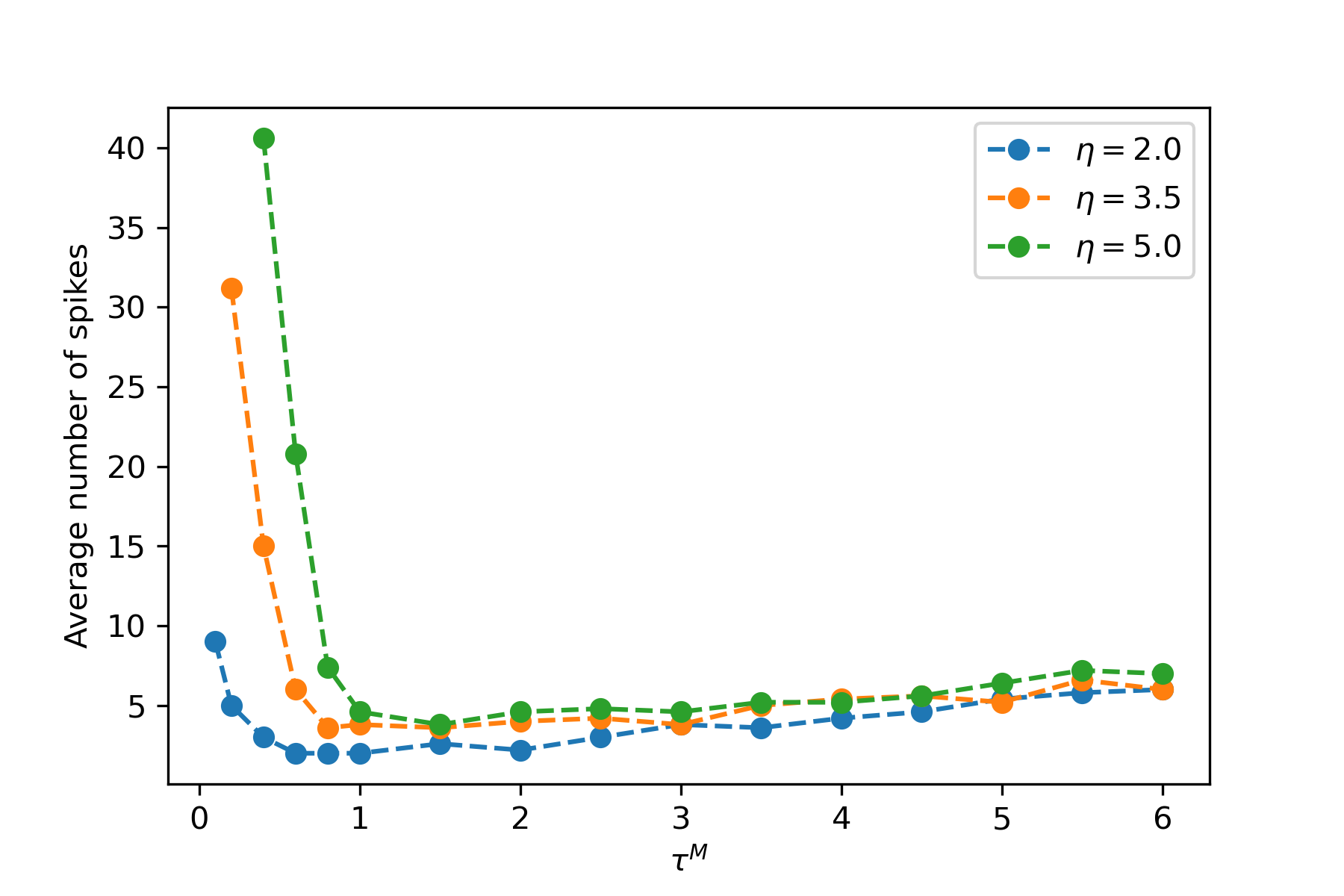}
  \caption{Under different values of $\eta$}\label{subfig:avenumspeta}
 \end{subfigure}\\ 
\caption{Average number of spikes}\label{fig:avenumsp}
\end{figure}

Let us see the relation between the parameters and the number of the spikes. For each combination of the parameters, five simulations were performed. The initial values for each simulation were generated randomly, therefore the initial values generally differ for each simulation. The line graphs in Figure \ref{fig:avenumsp} display the average number of spikes obtained in the five simulations for $\tau^M=6.0$, $5.5$, $5.0$, $4.5$, $4.0$, $3.5$, $3.0$, $2.5$, $2.0$, $1.5$, $1.0$, $0.8$, $0.6$, $0.4$, $0.2$, and $0.1$.\footnote{
The parameters are set as follows.\\
Fig. \ref{fig:avenumsp}\subref{subfig:avenumspta}: $\rho=1.0$, $\mu=0.5$, $\sigma=3.0$, and $\eta=2.0$.\\
Fig. \ref{fig:avenumsp}\subref{subfig:avenumspeta}: $\rho=1.0$, $\mu=0.5$, $\sigma=3.0$, and $\tau^A=2.0$.\\
In these figures, the actual computed values are indicated by the dots. The dashed lines are interpolation for the plot.} We can see a redispersion tendency for the average number of spikes to initially decreases and then increases again as $\tau^M$ value decreases. Figures \ref{fig:avenumsp}\subref{subfig:avenumspta} and \ref{fig:avenumsp}\subref{subfig:avenumspeta} show how the line graph change as $\tau^A$ and $\eta$ change, respectively. Now, we can see that the line graph tends to shift downward as the value of $\tau^A$ and $\eta$ decrease. This implies that lower agricultural transportat costs and stronger preference for agricultural variety promote agglomeration as a reduction in the average number of spikes.

\section{Conclusion}

We have considered the CP model with transport costs of differentiated agricultural goods on a continuous space. In particular, we have investigated the stability of the homogeneous stationary solution and the asymptotic behavior of time-evolving solutions to the model on the one-dimensional periodic space.

Each eigenfunction is found to behave similarly to the solution around the homogeneous stationary solution of the two-regional model, i.e., it is stable at sufficiently high manufacturing transport costs, unstable at moderate manufacturing transport costs, and stable again at low manufacturing transport costs as seen in Figure \ref{eigsmore}. This shows that the redispersion is also observed in the continuous space model.

The unstable domains are numerically observed to expand in general as the frequency increases as seen in Figures \ref{fig:ta_ccs} and \ref{fig:eta_ccs}. This implies that lower agricultural transportat costs and stronger preference for agricultural variety tend to promote agglomeration.

It is numerically observed that solutions starting around the homogeneous stationary solution asymptotically converge to spatial distributions having several spikes as seen in Figures \ref{fig:stsol1} to \ref{fig:stsol3}. In particular, from Figure \ref{fig:avenumsp}, we can see the redispersion in terms of the average number of the spikes. That is, as manufacturing transport costs decrease, the number of the spikes goes from decreasing to increasing. Furthermore, lower agricultural transport costs and stronger preferences for agricultural variety are suggested to promote agglomeration by reducing the average number of the spikes.

\section{Appendix}

\subsection{Proof for Theorem \ref{th: stationary solution}}\label{proof:thstsol}

From \eqref{totlam1} and \eqref{racetrack integration}, $\ol{\lambda}$ must be
\[
\begin{aligned}
&\ol{\lambda} = \frac{1}{2\pi \rho}.
\end{aligned}
\]
From the first equation of \eqref{consys}, $\ol{Y}$ satisfies
\begin{equation}\label{olY}
\ol{Y}=\mu\ol{w}^{M}\ol{\lambda} + (1-\mu)\ol{w}^{A}\ol{\phi}.
\end{equation}
From the second equation of \eqref{consys}, $\ol{G}^{A}$ becomes
\begin{equation}\label{olGA}
\begin{aligned}
\ol{G}^{A}
&= \left[\ol{\phi}{\ol{w}^{A}}^{1-\eta}
E_\alpha\right]^{\frac{1}{1-\eta}}.
\end{aligned}
\end{equation}
From the third equation of \eqref{consys}, we have
\begin{equation}\label{olw}
\begin{aligned}
\ol{w}^{A} 
&= \left[\ol{Y}{\ol{G}^{A}}^{\eta-1} E_\alpha\right]^{\frac{1}{\eta}}.
\end{aligned}
\end{equation}
Applying \eqref{olY} and \eqref{olGA} to \eqref{olw} yields
\[
{\ol{w}^{A}} = {\ol{w}^{M}} (=:\ol{w}).
\]
This gives, from \eqref{olY}, 
\[
\ol{Y} = \frac{\ol{w}}{2\pi \rho}.
\]
From the fourth equation of \eqref{consys}, $\ol{G}^{M}$ becomes
\begin{equation}\label{olGM}
\begin{aligned}
{\ol{G}^{M}}
&= \left[\ol{\lambda} \ol{w}^{1-\sigma} E_\beta\right]^{\frac{1}{1-\sigma}}.
\end{aligned}
\end{equation}
From the sixth equation of \eqref{consys}, together with \eqref{olGA}, \eqref{olw} and \eqref{olGM}, homogeneous $\omega$ denoted by $\ol{\omega}$ is given by
\[
\begin{aligned}
\ol{\omega}
&= \ol{\lambda}^{\frac{\mu}{\sigma-1}}\ol{\phi}^{\frac{1-\mu}{\eta-1}}
E_\alpha^{\frac{1-\mu}{\eta-1}} E_\beta^{\frac{\mu}{\sigma-1}}.
\end{aligned}
\]
\qed

\subsection{Proof for Lemma \ref{Hs}}\label{proof:Hs}
Let $n$ be even. We see that
\[
H^\beta_n =
\frac{\beta^2r^2}{n^2+\beta^2r^2}.
\]
Putting $\beta r=X$, we have
\[
H^\beta_n 
=\frac{X^2}{n^2+X^2}.
\]
It is easy to see that
\[
\begin{aligned}
\lim_{X\to 0} \frac{X^2}{n^2+X^2} = 0,\\
\lim_{X\to \infty}\frac{X^2}{n^2+X^2} = 1.
\end{aligned}
\]
We also easily see that
\[
\frac{d}{dX}H^\beta_n = \frac{2n^2X}{(n^2+X^2)^2} > 0.
\]

Let $n$ be odd. In this case, we see
\[
H^\beta_n =
\frac{X^2(1+e^{-X\pi})}{(n^2+X^2)(1-e^{-X\pi})}
\]
by putting $X=\beta r$. By the L'Hopital's theorem, we have
\[
\lim_{X\to 0}H^\beta_n =
\lim_{X\to 0}\frac{2X(1+e^{-\pi X})+X^2(-\pi e^{-\pi X})}{2X(1-e^{-\pi X})+(n^2+X^2)\pi e^{-\pi X}} 
= 0.
\]
On the other hand, it is obvious that
\[
\lim_{X\to \infty}H^\beta_n =1.
\]
By differentiating $H^\beta_n$ by $X$, we obtain
\begin{equation}\label{ddXHbetanodd}
\frac{d}{dX}H^\beta_n 
= \frac{2n^2X(1-e^{-2\pi X})-2\pi X^2e^{-\pi X}(n^2+X^2)}{(n^2+X^2)^2(1-e^{-\pi X})^2}.
\end{equation}
To show the monotonicity of $H^\beta_n$ as for $X$, we need to prove that the numerator of \eqref{ddXHbetanodd} is positive. It is equivalent to 
\begin{equation}\label{epiXeminuspiXgt}
e^{\pi X}-e^{-\pi X} > \frac{\pi X}{n^2}(n^2+X^2).
\end{equation}
The left hand side of \eqref{epiXeminuspiXgt} is expanded into a series
\[
2\pi X +2\frac{(\pi X)^3}{3!}+2\frac{(\pi X)^5}{5!}+\cdots.
\]
Therefore, subtracting the right-hand side of \eqref{epiXeminuspiXgt} from the left-hand side of \eqref{epiXeminuspiXgt} yields
\begin{equation}\label{piX2piX33!}
\pi X+\left\{2\frac{(\pi X)^3}{3!}-\frac{\pi}{n^2}X^3\right\}+2\frac{(\pi X)^5}{5!}+\cdots.
\end{equation}
We are left with the task of proving the content of the braces in \eqref{piX2piX33!} is positive. It is sufficient to prove for $n=1$, and it is obvious that
\[
\pi\left(\frac{\pi^2}{3}-1\right)X^3>0.
\]
\qed

\subsection{Proof for Lemma \ref{mub1}}\label{proof:mub1}

It is easy to check that
\[
\frac{d}{dH^\alpha_n}b
= -(1-\mu)\frac{\eta+(\eta-1){H^\alpha_n}^2}{\left[\eta-(\eta-1){H^\alpha_n}^2\right]^2} < 0.
\]
It is also immediate that
\[
b=1 
\]
for $\alpha=0$ because $\alpha=0\Leftrightarrow H^\alpha_n=0$, and
\[
\lim_{H^\alpha_n \to 1} b = 1-\frac{1-\mu}{\eta-(\eta-1)}=\mu.
\]
\qed

\subsection{Proof for Lemma \ref{Dgt0}}\label{proof:Dgt0}
It is easy from Lemmas \ref{Hs} and \ref{mub1} to check
\[
D>\sigma-\frac{\mu}{\mu}\cdot1 -(\sigma-1)\cdot1^2=0.
\]
\qed

\subsection{Proof for Lemma \ref{Bgt0}}\label{proof:Bgt0}

It is evident from \eqref{B} and Lemma \ref{mub1}.
\qed
\vspace{5mm}

\subsection{Proof for Lemma \ref{lem: delomega}}\label{proof:delomega}

From \eqref{hatomegaOmeganhatlambda}, we have
\begin{equation}
\begin{aligned}
\varDelta\omega(t, x) =\varDelta\omega(t, \theta) &= \sum_{n=\pm1,\pm2,\cdots}\hat{\omega}_ne^{in\theta} \\
&=\sum_{n=\pm1,\pm2,\cdots}\Omega_n\hat{\lambda}_ne^{in\theta}.
\end{aligned}
\end{equation}
Therefore, 
\[
\int_S \varDelta\omega(t, x)dx = 
\sum_{n=\pm1,\pm2,\cdots}\Omega_n\hat{\lambda}_n
\rho\int_{-\pi}^{\pi} e^{in\theta} d\theta.
\]
Then, the fact that
\[
\int_{-\pi}^{\pi} e^{in\theta} d\theta = 0
\]
for $n=\pm1, \pm2, \cdots$ completes the proof.
\qed

\subsection{Proof for Theorem \ref{Th:NBH}}\label{proofNBH}

When the manufacturing transport costs go to infinity, i.e., when $\beta$ goes to infinity, the content in the square brackets of \eqref{omhat} becomes
\begin{equation}\label{1plusminsigmu2}
1+\frac{-\sigma\mu^2-\sigma b+\mu\sigma b+\mu\sigma-\mu}{b}.
\end{equation}
For there to be no black hole, \eqref{1plusminsigmu2} must be negative, for which it is sufficient that
\[
(1-\sigma+\mu\sigma)(b-\mu)<0.
\]
Since $b\in(\mu, 1]$ from Lemma \ref{mub1}, it immediately yields \eqref{nbh}.

\qed

\subsection{Proof for Theorem \ref{th:betatozero}}\label{betatozero}
When $\tau_M\to 0$ or $\sigma\to 1$, i.e., $\beta\to 0$, we see that $H^\beta_n \to 0$ by Lemma \ref{Hs}. Then, it is easy to see $D\to \sigma$ by \eqref{D}. As a result, it immediately follows from \eqref{ev} that 
\[
\Omega_n \to -\frac{2\pi\rho\ol{\omega}}{(\sigma-1)\sigma}B.
\]
It is obvious that this limit is negative since $B>0$ for $\alpha >0$ from Lemma \ref{Bgt0}.

\qed

\subsection{Proof for Theorem \ref{th:pev}}\label{proofpev}
From \eqref{def:Halphan}, we see that
\begin{equation}
\lim_{|n|\to\infty} H^\alpha_n = 0,\label{limnHalphan}
\end{equation}
Applying \eqref{limnHalphan} to \eqref{b} immediately yields
\begin{equation}\label{limnb}
\lim_{|n|\to\infty}b = 1.
\end{equation}
Applying \eqref{limnHalphan}, \eqref{limnb} to \eqref{B}, we see at once that
\begin{equation}\label{limnB}
\lim_{|n|\to\infty}B = 0.
\end{equation}
Since $D>0$ from Lemma \ref{Dgt0}, the sign of $\Omega_n$ is determined by that of $Q(H^\beta_n)$ where $Q$ is defined by \eqref{QH}. From \eqref{limnb} and \eqref{limnB}, we see that 
\begin{equation}\label{limnquadfun}
Q(H^\beta_n)=-\left[\sigma(1+\mu^2)-1\right]{H^\beta_n}^2 + \mu(2\sigma-1)H^\beta_n
\end{equation}
as $|n|\to\infty$. It is evident that $Q(H^\beta_n)>0$ for $H^\beta_n$ such that
\begin{equation}\label{0Hmu2sigma}
0 < H^\beta_n < \frac{\mu(2\sigma-1)}{\sigma(1+\mu^2)-1} ~(< 1).
\end{equation}
For any given $\beta>0$, $H^\beta_n$ can be made as close to $0$ as possible by making $|n|$ sufficiently large, and thus \eqref{0Hmu2sigma} holds.
\qed

\subsection{Proof for Theorem \ref{th:Omegatozero}}\label{Omegatozero}
From \eqref{def:Hbetan}, we see that
\begin{equation}
\lim_{|n|\to\infty} H^\beta_n = 0.\label{limnHbetan}
\end{equation}
Applying \eqref{limnHalphan}, \eqref{limnb}, \eqref{limnHbetan} to \eqref{D}, 
\begin{equation}\label{limnD}
\lim_{|n|\to\infty}D = \sigma. 
\end{equation}
As a result of \eqref{limnb}, \eqref{limnB}, \eqref{limnHbetan}, and \eqref{limnD}, we see from \eqref{omhat} that
\[
\lim_{|n|\to\infty} \Omega_n = 0.
\]

\qed

\subsection{Proof for Theorem \ref{th:monoexpalpha=0}}\label{proof:th:monoexpalpha=0}
The critical point that satisfies $\Omega_n=0$ is given by $\tau_M$ such that $H^\beta_n=H^*$ that is a solution of the quadratic equation $Q(H)=0$, where $Q(H)$ is given by \eqref{QH}. When $\alpha=0$, by \eqref{b} and \eqref{B}, we see that
\[
Q(H) = \left\{-\sigma\left(1+\mu^2\right)+1\right\}{H}^2
+\mu\left(2\sigma-1\right)H
\]
Therefore, the solutions of $Q(H)=0$ when $\alpha=0$ are zero and
\begin{equation}\label{rootsHbeta}
H^* = \frac{\mu(2\sigma-1)}{\sigma(1+\mu^2)-1}>0.
\end{equation}
By Lemma \ref{Hs}, we see that $H^\beta_n$ is monotonically increasing with respect to $\tau_M$ and that $\lim_{\beta\to 0}H^\beta_n=0$. Then, from \eqref{rootsHbeta}, the lower critical point for $n$ is $\tau_M$ satisfying $H^\beta_n = 0$, that is, $\tau_M = 0$. Meanwhile, the upper critical point for $n$ is $\tau_M$ such that $H^\beta_n=H^*$ from \eqref{rootsHbeta}.

It follows from \eqref{def:Hbetan} that
\begin{equation}\label{Hn>Hn+2}
\begin{aligned}
&H^\beta_{|n|} - H^\beta_{|n|+2}\\
&= \beta^2\rho^2\frac{1-(-1)^{|n|} e^{-\beta \rho \pi}}{1-e^{-\beta \rho \pi}}
\left(\frac{1}{|n|^2+\beta^2\rho^2}-\frac{1}{(|n|+2)^2+\beta^2\rho^2}\right) > 0.
\end{aligned}
\end{equation}

Suppose now that $\tau_M$ is the upper critical point for $n$ which satisfies $H^\beta_n=H^*$, then we immediately see that
\begin{equation}\label{n+2ineq}
H^\beta_{|n|+2} < H^*
\end{equation}
holds by \eqref{Hn>Hn+2}. Since $H^\beta_n$ is monotonically increasing with respect to $\tau_M$ for any $n$ from Lemma \ref{Hs}, the upper critical point for $n+2$ (when $n>0$) or $n-2$ (when $n<0$) must be increased from the current value for the equality to hold in \eqref{n+2ineq}.
\qed

\subsection{Numerical scheme}\label{subsec:numsch}
We describe the numerical scheme used in the simulations in Section \ref{sec:num}.
\subsubsection{Instantaneous equilibrium}
To compute the time evolution of the function $\lambda$ in the system \eqref{consys}, the equations for $Y$, $G^A$, $w^A$, $G^M$, $w^M$, and $\omega^M$ must be solved for $\lambda$ at any given time $t\geq 0$. Following \citet[Section 5.2]{FujiKrugVenab}, let us call the solutions $Y$, $G^A$, $w^A$, $G^M$, $w^M$, and $\omega^M$ of the first six equations of \eqref{consys} for a given $\lambda$ an {\it instantaneous equilibrium}. As can be seen from \eqref{consys}, in fact, for a given $\lambda$, only $w^A$ and $w^M$ are truly unknown functions, and the other functions of the instantaneous equilibrium can be obtained straightforwardly from $w^A$, $w^M$, and $\lambda$.

Then, let us see that $w^A$ and $w^M$ can be obtained by solving a fixed point problem. Substituting the first and the second equation into the third one of \eqref{consys}, and defining $W^A:={w^A}^\eta$ and $W^M:={w^M}^\sigma$, we have
\[
W^A(t,x)=
\int_S \frac{\mu{W^M(t,y)}^{\frac{1}{\sigma}}\lambda(t,y)+(1-\mu){W^A(t,y)}^{\frac{1}{\eta}}\phi(y)}{\int_S\phi(z){W^A(t,z)}^{\frac{\eta-1}{\eta}}e^{-\alpha d(y,z)}dz}e^{-\alpha d(x,y)}dy.
\]
Similarly, substituting the first and the fourth equations into the fifth one of \eqref{consys}, we have
\[
W^M(t,x)=
\int_S \frac{\mu{W^M(t,y)}^{\frac{1}{\sigma}}\lambda(t,y)+(1-\mu){W^A(t,y)}^{\frac{1}{\eta}}\phi(y)}{\int_S\lambda(t,z){W^M(t,z)}^{\frac{\sigma-1}{\sigma}}e^{-\beta d(y,z)}dz}e^{-\beta d(x,y)}dy.
\]
Therefore, if we define operators $\cl{F}^A_\lambda$ and $\cl{F}^M_\lambda$ as
\[
\begin{aligned}
&\cl{F}^A_\lambda(W^A, W^M)(t,x) \\
&\hspace{5mm}:= \int_S \frac{\mu{W^M(t,y)}^{\frac{1}{\sigma}}\lambda(t,y)+(1-\mu){W^A(t,y)}^{\frac{1}{\eta}}\phi(y)}{\int_S\phi(z){W^A(t,z)}^{\frac{\eta-1}{\eta}}e^{-\alpha d(y,z)}dz}e^{-\alpha d(x,y)}dy
\end{aligned}
\]
and
\[
\begin{aligned}
&\cl{F}^M_\lambda(W^A, W^M)(t,x)\\
&\hspace{5mm}:=\int_S \frac{\mu{W^M(t,y)}^{\frac{1}{\sigma}}\lambda(t,y)+(1-\mu){W^A(t,y)}^{\frac{1}{\eta}}\phi(y)}{\int_S\lambda(t,z){W^M(t,z)}^{\frac{\sigma-1}{\sigma}}e^{-\beta d(y,z)}dz}e^{-\beta d(x,y)}dy,
\end{aligned}
\]
then we see that $(W^A, W^M)$ of the instantaneous equilibrium at any given time $t\geq 0$ is the fixed point\footnote{The idea of formulating an instantaneous equilibrium of the original CP model as such a fixed point problem is based on \citet{TabaEshiSakaTaka}.} satisfying  
\begin{equation}\label{fixedpoint}
\left[
\begin{array}{c}
W^A\\
W^M
\end{array}
\right]
=
\left[
\begin{array}{c}
\cl{F}^A_\lambda(W^A, W^M) \\[1mm]
\cl{F}^M_\lambda(W^A, W^M) 
\end{array}
\right].
\end{equation}
Finding a solution to the fixed point problem \eqref{fixedpoint} would be theoretically difficult, but in our model, it can be solved in most cases by a simple iterative method. That is, for a current solution candidate $(W^A_{\rm old}, W^M_{\rm old})$, the new solution candidate $(W^A_{\rm new}, W^M_{\rm new})$ is simply given by
\begin{equation}\label{iteration}
\left[
\begin{array}{c}
W^A_{\rm new}\\
W^M_{\rm new}
\end{array}
\right]
=
\left[
\begin{array}{c}
\cl{F}^A_\lambda(W^A_{\rm old}, W^M_{\rm old}) \\[1mm]
\cl{F}^M_\lambda(W^A_{\rm old}, W^M_{\rm old}) 
\end{array}
\right],
\end{equation}
and we iterate this until $\left\|W^A_{\rm new}-W^A_{\rm old}\right\|_\infty<10^{-10}$ and $\left\|W^M_{\rm new}-W^M_{\rm old}\right\|_\infty<10^{-10}$, with the final $(W^A_{\rm new}, W^M_{\rm new})$ being an approximated solution to \eqref{fixedpoint}.

In the actual numerical computation, the integration and the derivative must be discretized appropriately. The operators $\cl{F}^A_\lambda$ and $\cl{F}^M_\lambda$ are approximated by replacing the integral with a Riemann sum\footnote{Under the periodic boundary condition, the Riemann sum has the same accuracy as the trapezoidal rule.} as
\[
\cl{F}^A_\lambda(W^A, W^M)(t_n,x_i) \simeq
\sum_{j=1}^I \frac{\mu{{W^M}^n_j}^{\frac{1}{\sigma}}\lambda^n_j+(1-\mu){{W^A}^n_j}^{\frac{1}{\eta}}\phi_j}{\sum_{k=1}^I\phi_k{{W^A}^n_k}^{\frac{\eta-1}{\eta}}e^{-\alpha d(x_j,x_k)}\varDelta\theta}e^{-\alpha d(x_i,x_j)}\varDelta\theta
\]
and 
\[
\cl{F}^A_\lambda(W^A, W^M)(t_n,x_i) \simeq
\sum_{j=1}^I \frac{\mu{W^M_j}^{\frac{1}{\sigma}}\lambda^n_j+(1-\mu){W^A_j}^{\frac{1}{\eta}}\phi_j}{\sum_{k=1}^I\lambda^n_k{W^M_k}^{\frac{\sigma-1}{\sigma}}e^{-\beta d(x_j,x_k)}\varDelta\theta}e^{-\beta d(x_i,x_j)}\varDelta\theta
\]
respectively.\footnote{$\phi_k$, $k=1,2,\cdots,I$ are discretized values of $\phi(x)$.} Thus, once $W^A$ and $W^M$ are in hand, the approximated $w^A$,$w^M$, $G^A$, $G^M$, and $\omega^M$ can be computed as 
\[
\begin{aligned}
&{w^A}^n_i = {{W^A}^n_i}^{\frac{1}{\eta}},~{w^M}^n_i = {{W^M}^n_i}^{\frac{1}{\sigma}},\\
&{G^A}^n_i =
\left[\sum_{j=1}^I \phi_j{{w^{A}}^n_j}^{1-\eta} e^{-\alpha d(x_i,x_j)}\varDelta\theta\right]^{\frac{1}{1-\eta}},\\
&{G^M}^n_i = 
\left[\sum_{j=1}^I \lambda^n_j{{w^{M}}^n_j}^{1-\sigma} e^{-\beta d(x_i,x_j)}\varDelta\theta\right]^{\frac{1}{1-\sigma}},\\
&{\omega^M}^n_i = {w^M}^n_i{{G^M}^n_i}^{-\mu}{{G^A}^n_i}^{\mu-1}.
\end{aligned}
\]
The differential equation in \eqref{consys} is approximated by the Euler method as
\[
\lambda^{n+1}_i = \lambda^n_i + \varDelta t\left({\omega^M}^n_i - \sum_{j=1}^I{\omega^M}^n_j\lambda^n_j\varDelta\theta\right)\lambda^n_i.
\]

\vspace{10mm}
\noindent
{\bf Acknowledgements}

The author would like to express sincere gratitude to the anonymous reviewers. Their insightful and constructive comments have improved the paper. In particular, one of the reviewers suggested that \eqref{Hn>Hn+2} holds.

\bibliographystyle{aer}

\ifx\undefined\bysame
\newcommand{\bysame}{\leavevmode\hbox to\leftmargin{\hrulefill\,\,}}
\fi

\end{document}